\documentclass[fleqn]{article}
\usepackage[utf8]{inputenc}
\usepackage{amsmath}
\usepackage{xcolor}
\usepackage{graphicx}
\usepackage{tikz}
\usepackage{float}
\usepackage{amsfonts}
\usepackage{amssymb}
\usepackage[toc,page]{appendix}
\usepackage[top=1in,bottom=1in,right=1in,left=1in]{geometry}
\usepackage{amsthm}
\usepackage{subfiles}
\DeclareGraphicsExtensions{.pdf,.png}

\usetikzlibrary{positioning}

\title{Black hole microstructures in the extremal limit}

\author{George Ruppeiner\footnote{correspondence: ruppeiner@ncf.edu}\,\, and Alexandru-Mihail Sturzu\footnote{Division of Natural Sciences, New College of Florida, 5800 Bay Shore Road, Sarasota, FL 34243, USA}
}

\date{\today}

\begin{document}

\maketitle

\begin{abstract}
The microstructure of black holes is a mystery, with no general consensus on questions as basic as to what the constituent particles (if any) might be. We approach these questions with black hole thermodynamics (BHT), augmented with the metric geometry of thermodynamics. This geometry connects to interparticle interactions via the invariant thermodynamic Ricci scalar curvature $R$, which may be calculated with BHT. In ordinary thermodynamics (OT), $R$ is positive/negative for interparticle interactions that are repulsive/attractive. Its magnitude is the correlation length. The basic universality of thermodynamics leads us to expect similar relations for BHT. Our contribution here is motivated by a physical simplification that frequently occurs at low temperatures $T$ in OT: complicated interparticle interactions tend to freeze out, leaving only the basic quantum statistical interactions, those of ideal Fermi and Bose gasses. Our hope is that a similar simplification happens in black holes in the extremal limit, where the BHT temperature $T \to 0$. We evaluate the extremal regime for twelve BHT models from the literature, working with the independent variables mass, angular momentum, charge, and the cosmological constant, $\{M,J,Q,\Lambda\}$ respectively. We allowed only two of these variables to fluctuate at a time, with the other two fixed. $M$ always fluctuated, either $J$ or $Q$ fluctuated, and $\Lambda$ was always fixed. We found that, at constant average $M$, the thermodynamic invariant $R$ has a limiting divergence $R=c\,T^{-1}$, with the nonsingular constant $c$ depending only on $M$ and the two fixed parameters. $c$ is positive for $11/12$ of the models we examined, and negative only for the tidal charged model. The positive sign for $R$ indicates a BHT microstructure composed of particles with repulsive (fermionic) interactions. The limiting BHT expression for $R$ resembles that for the two and three-dimensional ideal Fermi gasses at constant volume, which also have a limiting divergence $R=c\,T^{-1}$, and with a positive $c$.

\end{abstract}


\section{Introduction}\label{S: Introduction}

What is the microstructure of black holes? This question is unsettled as theoretical difficulties and a lack of relevant experimental data have limited progress. Unclear at this point are properties as fundamental as what black holes are made of. Are black holes fundamentally macroscopic entities, with no microstructure at all, as the ``no-hair'' conjecture suggests? Or are they composed of some type of known or unknown microscopic particles or strings? If so, do these microscopic constituents fill the interior of the black hole volume, have they collapsed to the center, or are they concentrated at the event horizon? The continuing success of the General Theory of Relativity (GR) in explaining observational results certainly supports a macroscopic picture. But recent experimental efforts such as the detection of gravitational waves \cite{Abbott2016}, and images from the event horizon telescope \cite{Doeleman2019} offer real possibilities for results beyond the predictions of GR.

\par
In our paper we take a theoretical approach to black hole microstructures, one based on black hole thermodynamics (BHT) \cite{Bekenstein1980, Wald2001, Carroll2019}. In BHT, macroscopic black hole properties such as mass, angular momentum, and charge, $(M,J,Q)$ respectively, are elements of a structure that follows the laws of thermodynamics. Details (e.g., thermodynamic equations of state) can be found from the Bekenstein-Hawking relation defining the black hole entropy $S$ in terms of the area $A$ of the event horizon \cite{Bekenstein1973, Hawking1976}:

\begin{equation} 
\frac{S}{k_B}=\frac{1}{4}\left(\frac{A}{L_p^2}\right),
\label{10}
\end{equation}

\noindent where $k_B$ is Boltzmann's constant, and $L_p$ is the Plank length:

\begin{equation}L_p=\sqrt{\frac{\hbar G}{c^3}},\label{20}\end{equation}

\noindent with $\hbar$ the reduced Planck's constant, $G$ the gravitational constant, and $c$ the speed of light. Note that $\{M,J,Q\}$ are all conserved quantities.

\par
The calculation of $A=A(M,J,Q)$ with GR, or by other means, leads with Eq. (\ref{10}) to the fundamental thermodynamic equation $S=S(M,J,Q)$ providing all of the BHT properties, as we will explicitly demonstrate in Section \ref{S: Results}. Since $L_p$ contains Planck's constant, BHT naturally brings quantum mechanics into the purely classical GR regime, and thus offers at least some bare basics of a quantum-gravity picture.

\par
But what does BHT tell us about the microscopic elements of black holes? Key here is that any thermodynamic structure contains within it a fluctuation theory that links macroscopic properties to microscopic properties \cite{pathria2011}. To sort out what thermodynamic fluctuations are telling us about microscopic properties, we employ the thermodynamic Ricci curvature scalar $R$ \cite{Ruppeiner1979, Ruppeiner1995a}. 

\par
$R$ is a thermodynamic invariant in the metric geometry of thermodynamics. In ordinary thermodynamics (OT), the magnitude $|R|$ gives the average volume of groups of atoms organized by their interparticle interactions. Near critical points, this average volume is given by the correlation length. The sign of $R$ gives the basic character of the interparticle interactions: $R$ is positive/negative for interactions repulsive/attractive, in the curvature sign convention of Weinberg \cite{Weinberg1972}, in which $R$ for the two sphere is negative. If we believe in the basic universality of thermodynamics, we might expect these features of $R$ to apply in the black hole scenario as well. For a brief review of the geometry of thermodynamics in the integrated environment of OT and BHT, see \cite{Ruppeiner2014}. For discussion about restricted fluctuations, where one or more of $\{M,J,Q\}$ are fixed, see \cite{Ruppeiner2007,Sahay2017b}. Restricted fluctuations are important in Section \ref{S: Results}.

\par
In our paper we numerically calculate $R$ for a number of published BHT models that developed analytic equations for the quantities $S$, $T$, and $R$ in terms of $\{M, J, Q\}$ or $M$, $T$, and $R$ in terms of $\{S, J, Q\}$. In some of these models, the cosmological constant $\Lambda$ is included as a static parameter. In contrast to most evaluations of BHT models, which focus on Van der Waals type phase transitions at nonzero temperatures, our focus is on the black hole extremal limit where $T\to 0$.

\par
We argue that the extremal limit offers the possibility of a direct connection between the microstructures behind BHT, and those corresponding to OT. The reason for this is that in OT, as $T\to 0$, the physics frequently simplifies to its absolute basics, which is in many cases the elementary ideal Fermi or Bose gasses. This happens as the more complicated interactions between the constituent particles freeze out. For example, this is a reason why the ideal Fermi gas is so effective in leading to an understanding of the physical stability of white dwarfs \cite{Chandrasekhar1967}. White dwarfs are held up by free electrons, and have temperature of the order of less than the Fermi temperature. Note as well that in laboratories, both ideal Bose and Fermi gasses have been produced by cooling to micro Kelvin temperatures \cite{Jin1999}. The thermodynamics agrees remarkably well with the degenerate ideal gasses. Our hope is that this freezing out of the microstates occurs in black holes as well, and that the BHT in the extremal limit takes on the character of either an ideal Fermi or ideal Bose gas. We test this conjecture by comparing the thermodynamic invariant $R$ for BHT and the ideal Fermi gas, in the limit $T\to 0$.

\par
We find a measure of consistency between the limiting BHT and ideal Fermi cases. We calculated $R$ for twelve BHT models. In eleven of these models, $R$ diverges to positive infinity in the extremal limit. Furthermore, along curves of constant $M$, the divergence is as $R=c/T$, where $c$ is a constant depending only on $M$ and two fixed parameters $J$ or $Q$, and $\Lambda$. This dependence matches that of the ideal Fermi gas in 2D and 3D. In 2D, the proportionality constant $c$ is independent of the system mass, and in 3D it depends somewhat on the mass. However, we attempt no systematic comparison of the mass dependence of $c$ between BHT and the ideal Fermi gas. Such a project might better be done with a more sophisticated Fermi gas model than the one employed employed here. The lone exception to the extremal fermionic behaviour is the tidal charged model \cite{dadhich2000, Pido2011}. There, the divergence in $R$ is negative in the extremal limit, more similar to that of the Bose gas. We offer no explanation for this.

\par
The match between BHT models and the ideal Fermi gas in the extremal limit has been reported previously in the Kerr-Newman family of black holes \cite{Ruppeiner2008}, and we systematically extend it here to a number of various BHT models. But the proposition that the microstructure of black holes should be composed of fermions is not really surprising. The condensed matter white dwarfs are held up against gravity by a gas of largely free electrons, modeled as a Fermi ideal gas. The much more condensed neutron stars are supported against gravitational collapse by fermionic neutrons. It is hence a reasonable extrapolation that black holes, only a little denser than neutron stars, also be composed of fermions.\footnote{We acknowledge a webinar on itelescope on 4/7/2023 by Andrealuna Pizzetti for this elementary argument.} Our contribution in this paper is establishing a clear, systematic connection between fermionic properties and models of general relativity and string theory.

\par Table \ref{ResultsTable} lists the models we analyzed in this paper, and states some basic outcomes.

 \begin{table}
    \centering
    \begin{tabular}{lcccc}
    Black hole model & Params & Diverge & $R$-sign & Analytic \\
    \hline
    Kerr                        & $\{M, J\}$ & $T^{-1}$ & + & Yes \\
    Kerr-Newman (fixed $Q$)     & $\{M, J\}$ & $T^{-1}$ & + & Yes \\
    Kerr-Newman (fixed $J$)     & $\{M, Q\}$ & $T^{-1}$ & + & Yes \\
    Kerr 5D                     & $\{M, J\}$ & $T^{-1}$ & + & Yes \\
    Black ring                 & $\{M, J\}$ & $T^{-1}$ & + & Yes \\
    Reissner-Nordstr{\"om} AdS  & $\{M, Q\}$ & $T^{-1}$ & + & Yes \\
    $f(R)$ gravity              & $\{M, Q\}$ & $T^{-1}$ & + & Yes \\
    Tidal charged               & $\{M, Q\}$ & $T^{-1}$ & - & Yes \\
    Gauss-Bonnet AdS            & $\{M, Q\}$ & $T^{-1}$ & + & Yes \\
    Dyonic charged AdS          & $\{M, Q\}$ & $T^{-1}$ & + & Yes \\
    Einstein-Dilaton            & $\{M, Q\}$ & $T^{-1}$ & + & No  \\
    $R$-charged                 & $\{M, Q\}$ & $T^{-1}$ & + & Yes \\
    \hline
    \end{tabular}
    \caption{Summary results table. ``Params'' are the pair of fluctuating parameters,``Diverge'' is the extremal $T$ dependence of the thermodynamic curvature $R$, ``$R$-sign'' is the sign of $R$ in the extremal limit, with ``+'' denoting fermionic, and ``Analytic'' denotes whether or not $S$, $T$, and $R$ may be written analytically.}
    \label{ResultsTable}
    \end{table}

\section{Thermodynamic geometry}

Basic thermodynamic metric geometry for BHT has been described in detail elsewhere \cite{Ruppeiner2007,Aaman2003}, so we keep the present discussion short. Let $S$ be the black hole entropy given in Eq. (\ref{10}). The thermodynamic metric has local distances corresponding to fluctuation probabilities: the less the probability of a fluctuation between two states, the further apart they are \cite{Ruppeiner1979, Ruppeiner1995a}. In this paper, we consider only two-dimensional thermodynamic metric \cite{Ruppeiner1995a} geometries, for which the line element is

\begin{equation}
    d\ell^2=g_{\mu\nu}dX^\mu dX^\nu,\label{Rmetric}
\end{equation}

\noindent with the coordinates $X=\{X^1,X^2\}$ taken as $M$ and one of $\{J,Q\}$. The two parameters other than $X$ are held fixed, if they are present in the model.

\par
The entropy metric elements are

\begin{equation}
g_{\alpha\beta}=-\frac{\partial^2 S}{\partial X^\alpha\partial X^\beta}.\label{E: thermo metric}
\end{equation}

\noindent This metric form requires that we know $S=S(X)$. However, we frequently know instead the function $M=M(Y)$, where the two fluctuating coordinates $Y$ are $S$ and one of $\{J,Q\}$. In principle, we could algebraically solve $M=M(Y)$ to get $S=S(X)$, but a closed form solution is usually difficult to find.

\par
In cases where we know $M(Y)$, but not $S(X)$, it is advisable to start with the Weinhold energy version of the thermodynamic metric \cite{weinhold1976}:

\begin{equation}d\ell^2_W=g_{\mu\nu,W}dY^\mu dY^\nu,\label{Wmetric}\end{equation}

\noindent where the Weinhold metric elements are

\begin{equation}g_{\alpha\beta,W}=-\frac{\partial^2 M}{\partial Y^\alpha\partial Y^\beta}.\label{}\end{equation}

\noindent The entropy version of the metric in Eq. (\ref{Rmetric}) that we really need follows from the identity \cite{Salamon1984}:

\begin{equation}d\ell^2=\frac{1}{T}d\ell^2_W,\label{145987}\end{equation}

\noindent where the temperature $T$ is given by

\begin{equation} \label{E: Temperature definition}
T=\left(\frac{\partial M}{\partial S}\right)_{J,Q,\Lambda}.
\end{equation}

\par
Generally, the thermodynamic Ricci curvature scalar is given by \cite{Sokolnikoff1964,Ruppeiner2012}

\begin{equation} \label{RicciScalar}
\begin{array}{lr}
{\displaystyle R= -\frac{1}{\sqrt{g}} \left[ \frac{\partial}{\partial x^1}\left(\frac{g_{12}}{g_{11}\sqrt{g}}\frac{\partial g_{11}}{\partial x^2}-\frac{1}{\sqrt{g}}\frac{\partial g_{22}}{\partial x^1}\right) \right. } \\  \hspace{3.6cm} + {\displaystyle \left. \frac{\partial}{\partial x^2}\left(\frac{2}{\sqrt{g}} \frac{\partial g_{12}}{\partial x^1} -\frac{1}{\sqrt{g}}\frac{\partial g_{11}}{\partial x^2}-\frac{g_{12}}{g_{11}\sqrt{g}}\frac{\partial g_{11}}{\partial x^1}\right)\right],}
\end{array} 
\end{equation}

\noindent where

\begin{equation}
    g = \det{g_{\alpha \beta}} = g_{11} g_{22} - g_{12}^2.
\end{equation}

\section{Ideal Fermi gasses}

The basic properties of the ideal Fermi gas thermodynamics are well-known \cite{pathria2011}. This gas consists of $N$ non-interacting fermions, each with mass $m$, confined to a box with fixed size and hard walls. The potential is zero inside the box. We consider only boxes with two and three dimensions, corresponding to cases where the hypothetical BHT particles all reside on the event horizon or fill the full volume inside the event horizon.

\subsection{2D ideal Fermi gas}

The thermodynamic curvature $R$ for the 2D ideal Fermi gas was worked out in \cite{Ruppeiner2008}. In the Thomas-Fermi continuum approximation, the thermodynamic potential per area $\phi=p/T$ can be expressed as

\begin{equation} \phi\left(\frac{1}{T},-\frac{\mu}{T}\right) = k_B (2s+1)\lambda^{-2}f_2(\eta), \label{3758399012} \end{equation}

\noindent with pressure $p$, fugacity $\eta=\mbox{exp}(\mu/k_B T)$, chemical potential $\mu$, thermal wavelength

\begin{equation} \lambda = \frac{h}{\sqrt{2\pi m k_B T}}, \label{} \end{equation}

\noindent particle spin $s$, Planck's constant $h$, particle mass $m$, and

\begin{equation}f_l(\eta)=-\mbox{PolyLog}(l,-\eta). \end{equation}
 
\noindent The function $\phi$ in Eq. (\ref{3758399012}) is naturally written in the coordinates $\{F_1,F_2\}=\{1/T,-\mu/T\}$. $\phi= \phi(F_1,F_2)$ yields all of the thermodynamics.

\par
The energy and particle number, both per area, are $\{u,\rho\}=\{-\phi_{,F_1},-\phi_{,F_2}\}$, where the comma notation indicates partial differentiation. We have

\begin{equation} u=(2s+1)\,k_B T \,\lambda^{-2}\, f_2(\eta), \end{equation}

\noindent and

\begin{equation}\rho= (2s+1)\lambda^{-2} \ln(1+\eta). \label{653209} \end{equation}

\noindent In $\{F_1,F_2\}$ coordinates, the thermodynamic metric elements are \cite{Ruppeiner1995a}

\begin{equation}g_{\alpha\beta}=\frac{1}{k_B}\phi_{,\alpha\beta},\label{}\end{equation}

\noindent and the thermodynamic scalar curvature follows from Eq. (\ref{RicciScalar}):

\begin{equation} R=\lambda^2\,\frac{\eta\left[-(1+\eta)\ln^2(1+\eta)+(2\eta-\ln(\eta +1))f_2(\eta)\right]}{\left[(1+\eta) \ln^2(1+\eta)-2\eta f_2(\eta)\right]^2}.\label{103985} \end{equation}

\noindent Numerical calculations over a large grid of points indicate that in the physical range $-\infty<\mu<+\infty$, $0<T<\infty$, and $0<\eta<\infty$, $u$, $\rho$, and $R$ are all always positive.

\par
The low temperature comparison between the BHT models and the ideal Fermi gas must be done systematically. Generally, for functions of two variables, with one variable being taken to some limit, what is done with the other variable as we take the limit must be specified. But which other variable? For BHT, as $T\to 0$ we will always fix the total mass $M$, guided by the Kerr-Newman examples where the $R\propto 1/T$ result holds only with this variable fixed. For comparison with the ideal Fermi gas, we then likewise take the limit $T\to 0$ at fixed mass, or fixed number density $\rho$. We also looked at fixing the energy density $u$ for the ideal Fermi gas, but this too yields $R\propto +1/T$.

\par
From Eq. (\ref{653209}) we see that at fixed $T$, the density $\rho$ is an increasing function of $\eta$. For large $\eta$, the asymptotic expression becomes,

\begin{equation}\rho\to\frac{2\pi (2s+1)m}{h^2}\,\mu,\label{65743}\end{equation}

\noindent independent of $T$. 

\par
Now consider the regime of small $T$, and let $\rho$ be fixed at some value. Then, by Eq. (\ref{65743}), $\mu$ will likewise be fixed, and positive, leading to $\eta\to\infty$ as $T\to 0$. Very useful for dealing with the PolyLog function for large $\eta$ is the Sommerfeld approximation \cite{pathria2011}

\begin{equation}\text{PolyLog}(\nu,-e^{\xi}) = -\frac{\xi^{\nu}}{\Gamma (\nu +1)}\left[1+\nu(\nu -1)\,\frac{\pi^2}{6}\,\frac{1}{\xi^2}+O\left(\frac{1}{\xi^4}\right)\right].\label{}\end{equation}

\noindent Using this approximation in Eq. (\ref{103985}) yields the formula for small $T$:

\begin{equation} R \to \frac{3h^2}{2\pi^3 g m k_B T}, \label{limitingline1}\end{equation}

\noindent depending only on $T$, and not on $\rho$. For a check, the Mathematica ``Limit'' operation gives the same result. We also get the limiting expression for $u$

\begin{equation}u\to\frac{\pi (2 s+1)m}{h^2}\mu ^2,\label{}\end{equation}

\noindent which shows that lines of constant $u$ coincide with those of constant $\mu$, and thus constant $\rho$.

\par
\begin{figure} 
\begin{tabular}{c c}
        \includegraphics[width = 0.47\textwidth]{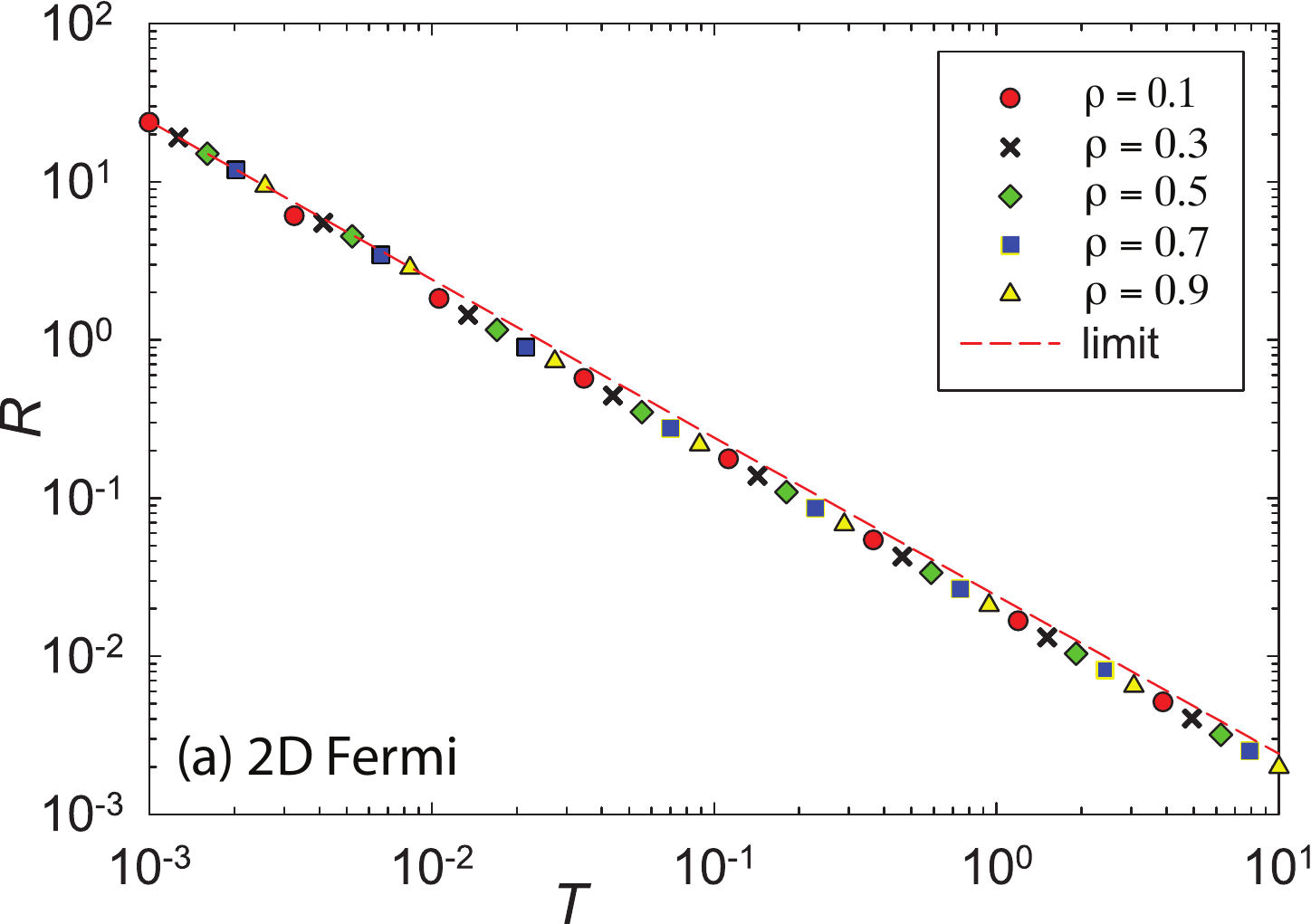} &
        \includegraphics[width = 0.46\textwidth]{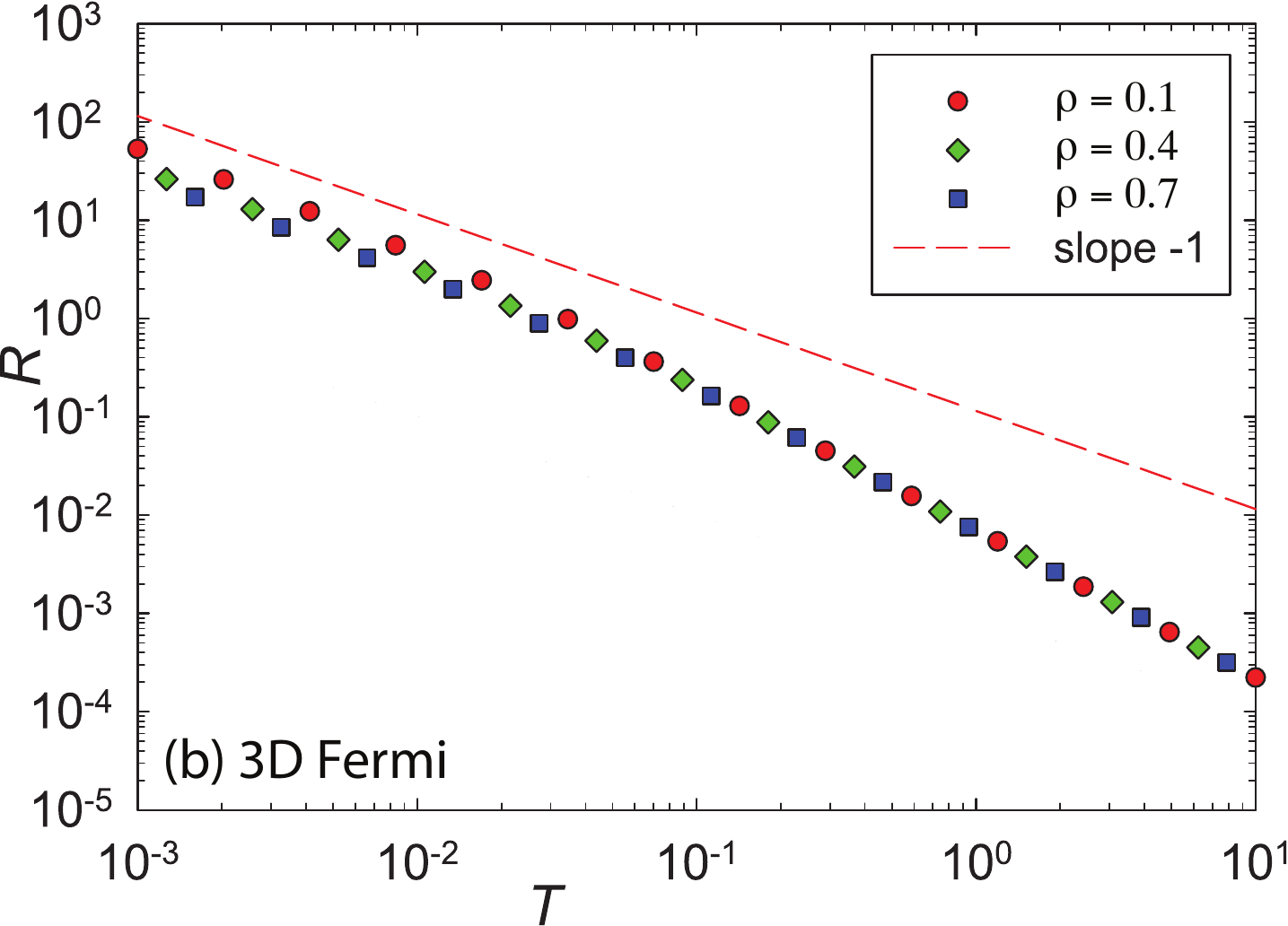} 
\end{tabular}
\caption{$R$ versus $T$ for the two and three-dimensional ideal Fermi gasses, each for several constant values of $\rho$. (a) For small $T$, the 2D case has all the points fall on the same curve, with limiting slope $-1$. (b) For small $T$, the 3D case shows a density dependence, but for each density the limiting points all fall on a curve with slope $-1$.}
\label{FigureFermi}
\end{figure}

Figure \ref{FigureFermi}(a) shows $R$ as a function of $T$ along several curves with constant $\rho$. Dimensionless units, with $k_B=1$, $h=1$, $m=1$, and $s=1/2$ have been used throughout the figures. The straight red line shows the asymptotic limiting expression in Eq. (\ref{limitingline1}). This line, independent of the value of $\rho$, clearly agrees with all of our values for $R$ at small $T$, as expected. Fair agreement with the asymptotic line extends into the larger temperatures regime. The data in the graph have $\eta$ ranging from $\sim 10^{-3}$ to $\sim 10^{34}$.

\subsection{3D ideal Fermi gas}

\par
The thermodynamic curvature $R$ for the 3D ideal Fermi gas was worked out in several places \cite{Janyszek1990,Oshima1999,Pessoa2021}. The thermodynamic potential per volume is \cite{pathria2011}

\begin{equation} \phi\left(\frac{1}{T},-\frac{\mu}{T}\right)=k_B (2s+1)\,\lambda^{-3}\, f_{\frac{5}{2}}(\eta).\label{65784} \end{equation}

\noindent This leads to

\begin{equation} u=\frac{3}{2}(2s+1)\lambda^{-3}\,f_{\frac{5}{2}}(\eta),\end{equation}

\begin{equation}\rho=(2 s+1)\lambda^{-3}\,f_{\frac{3}{2}}(\eta),\end{equation}

\noindent and

\begin{equation}R=5 \lambda ^3 \frac{\left[2 f_{\frac{5}{2}}\left(\eta \right)
   f_{\frac{1}{2}}\left(\eta \right)^2-f_{\frac{3}{2}}\left(\eta \right)^2
   f_{\frac{1}{2}}\left(\eta \right)-f_{-\frac{1}{2}}\left(\eta \right)
   f_{\frac{3}{2}}\left(\eta \right) f_{\frac{5}{2}}\left(\eta
   \right)\right]}{(2 s+1) \left[3 f_{\frac{3}{2}}\left(\eta \right)^2-5
   f_{\frac{1}{2}}\left(\eta \right) f_{\frac{5}{2}}\left(\eta
   \right)\right]^2}.\end{equation}

\noindent This equation for $R$ matches Eq. (13) in ref. \cite{Oshima1999}. (There is a small error in the corresponding Eq. (4.21) of ref. \cite{Janyszek1990}.) Over the physical range $-\infty<\mu<+\infty$, $0<T<\infty$, and $0<\eta<\infty$, $u$, $\rho$, and $R$ are all always positive.

\par
For large $\eta$, the Sommerfeld approximation yields

\begin{equation}\rho\to\frac{8\sqrt{2}\,\pi (2s+1) m^{3/2}}{3 h^3}\mu^{3/2},\label{}\end{equation}

\noindent independent of $T$. Along curves of constant $\rho$, $\mu$ stays fixed, and $\eta\to\infty$ as $T\to 0$, self-consistent with our use of the Sommerfeld approximation above. We also have the limiting form

\begin{equation}u\to \frac{8 \sqrt{2}\,\pi (2 s+1) m^{3/2} }{5 h^3}\mu ^{5/2},\label{}\end{equation}

\noindent which shows that, at low $T$, lines of constant $u$ correspond to lines of constant $\rho$. Finally, the limiting asymptotic form for $R$ is:

\begin{equation}R\to \frac{3^{2/3} h^2 \sqrt[6]{512 s^2+512 s+128}}{4 \sqrt{2} \pi^{8/3} m (2s+1)}\left(\frac{1}{k_B T \rho^{1/3}}\right),\label{limitingline2}\end{equation}

\noindent matching the $R\propto +1/T$ form present in the 2D ideal Fermi gas. The 3D case does, however, have a mild dependence on $\rho$ in contrast to that for 2D, which has none.

\par
Our finding $R\propto +1/T$ for the ideal Fermi gasses matches the divergences in the extremal limits for 11/12 of the BHT models we consider below. This internal consistency among the BHT results, as well as with the ideal Fermi gas divergences, is the main result in our paper.

\par
But further work might produce improved results. First, in an earlier work \cite{Ruppeiner2008} it was reported incorrectly that the 3D ideal Fermi gas had limiting divergence $R\propto+1/T^{3/2}$. This led to a claim that a filling of the full volume of the black hole interior was inconsistent with the BHT Kerr-Newman models considered in that reference. It was suggested instead that the Fermi gases resides on the event horizon itself, since the 2D ideal Fermi gas has the $R\propto+1/T$ divergence. This idea is flawed, however, since our more accurate calculation for the 3D ideal Fermi gas presented here, with its $R\propto+1/T$  divergence, leaves a claim of how the particles constituting the interior of the black hole arrange themselves at best premature. Second, as we discuss below, the constants of proportionality multiplying $1/T$ are at present difficult to match between ideal Fermi and BHT.

\par
The task of sorting these issues out might be greatly assisted by the introduction of superior ideal Fermi gas models. Our choice here was to pick the simplest models, but perhaps more creative models might be viable. But this issue is beyond the scope of this paper.

\section{Research protocol} \label{S: Research Protocol}

\par
There is a large literature on probing black hole microstructures with the thermodynamic curvature and a systematic selection among evaluated BHT models was necessary to keep our project manageable. Many of the evaluated BHT models are based on combinations of parameters chosen from among four fundamental fluctuating quantities: $\{M, J, Q, \Lambda\}$, where $\Lambda$ is the cosmological constant. This quartet of values is enough to specify the BHT state for all of the models that we considered.

\par
We considered only models where the thermodynamic geometry is two-dimensional. So only two of $\{M, J, Q, \Lambda\}$ fluctuate while the other two are fixed. Shen et al. \cite{Shen2007} used the Legendre transformed quantity $M-\phi Q$, where $\phi$ is the electrostatic potential on the event horizon, in place of $M$ in the thermodynamic metric. But such approaches are beyond the scope of this manuscript. We also did not evaluate cases of ``extended thermodynamics,'' where $\Lambda$ is a fluctuating thermodynamic parameter. In extended thermodynamics, $\Lambda$ connects to the pressure, conjugate to the black hole volume. Much recent work has been done here, see e.g. \cite{Wei2019}, and we leave the project of sorting out the extended thermodynamic models in the extremal limit to more qualified authors. For us, $\Lambda$ was always fixed. We also restrict ourselves to cases where the Bekenstein-Hawking equation, Eq. (\ref{10}), holds exactly, without the logarithmic correction terms occasionally seen.

\par
Calculations for the BHT models we consulted in the literature were frequently quite involved, and we made no systematic attempt to verify them for correctness. What we needed from each model were analytic equations for the functions $S$, $T$, and $R$ in terms of $\{M, J, Q,\Lambda\}$ or $M$, $T$, and $R$ in terms of $\{S, J, Q,\Lambda\}$. $R$ could be positive or negative depending on the curvature sign convention. The sign convention employed was usually clear in each paper. We expressed all of the $R$'s in this paper in our curvature sign convention (i.e. fermionic has positive $R$). Harder to sort out are the systems of units employed in the literature, and we felt that it would be too confusing (and even prone to error) to impose our own uniform unit system here. This means that our graphs of $R=R(M,T)$ were not necessarily consistent across different models, differing by scaling factors.

\par
Our approach was numerical, and centered around the three basic functions for $S$ (or $M$), $T$, and $R$. Other than these three functions from the BHT models, our analysis is independent of the specifics of the BHT models. Such simplification is essential in order to handle a number of disparate models effectively. Details of our coding algorithm may be found in the Appendix.

\section{Results} \label{S: Results}

In this section we discuss the results for the thermodynamic scalar curvature $R$ for twelve different BHT models. The models were selected according to the criteria established in Section 4, and all have two-parameter BHT's, with either fluctuating $\{M,J\}$ or fluctuating $\{M,Q\}$. Thermodynamic stability, i.e., a positive definite thermodynamic metric, is a somewhat mixed proposition in BHT. We attempt no systematic stability analysis here. But we do pass along stability results reported in the literature.

\par
\begin{figure}[h!]
    \centering
        \includegraphics[width = 0.5\textwidth]{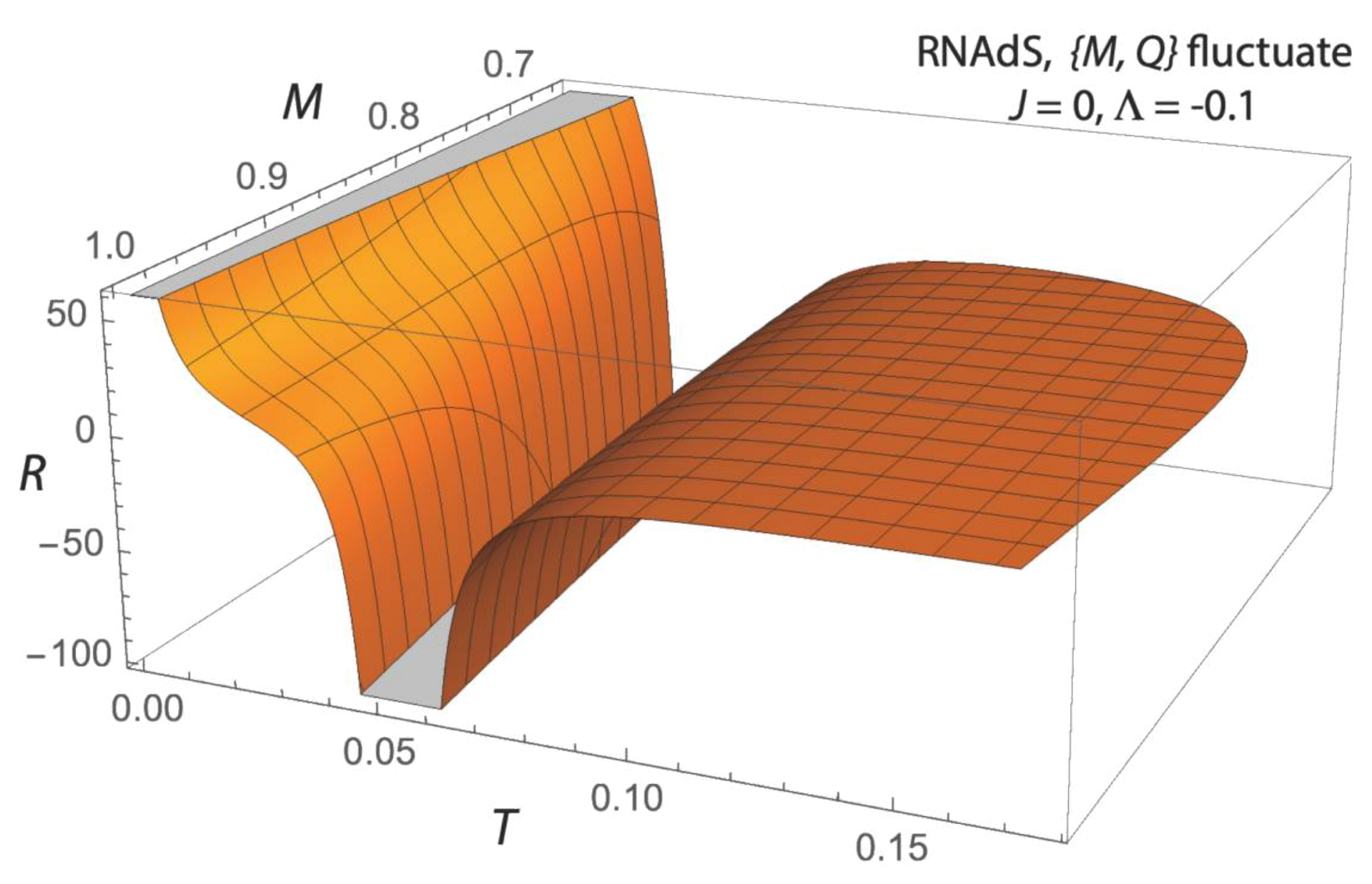}
        \caption{$R$ versus $(M,T)$ for RNAdS with $J=0$ and $\Lambda=-0.1$. We see three regimes of interest: 1) a low temperature extremal limit regime with positive $R$, 2) an intermediate temperature regime with a line of phase transitions indicated by $R$ diverging to minus infinity, and 3) a high temperature ideal gas regime with $|R|$ near zero.}
    \label{fig:3Dplot}
\end{figure}

Before considering the individual cases in detail, we start with a simple graph, offered by the RNAdS BHT model worked out by {\AA}man et al. \cite{Aaman2003}, and discussed in detail here in subsection 5.6. Figure \ref{fig:3Dplot} shows a plot for $R$ as a function of $(M,T)$, with $J=0$ and $\Lambda=-0.1$. The corresponding contour plot is shown in Figure \ref{F: 3D Plots}(f).

\par
$R$ diverges to plus infinity as $T$ decreases to zero. For the values of $M$ represented here, there is a line of phase transitions where $R$ diverges to minus infinity. Such lines of divergence signal second-order phase transitions, generally of great interest in the BHT literature. $|R|$ gets small for $T$ above the phase transition. In the picture that we develop here, this is because the particles constituting the microstructure have effective interparticle interactions less strong with increasing temperature. For a given $M$ value, the contour surface eventually terminates as $T$ reaches a maximum value, as discussed in subsection 5.1.

\par The basic parallel between this black hole figure and those for examples in OT is remarkable.

\subsection{Kerr $\{M,\,J\}$} \label{S: Kerr}

We start with the Kerr black hole, because it is well-known, is relatively simple, and has physical relevance. Kerr also has a known analytic expression for $R$ near the extremal limit, and this provides guidance for our entire project. To set some of our themes, we spend a bit more time on its presentation. Kerr black holes are spinning, uncharged systems with $\{Q,\Lambda\}=\{0,0\}$, and with BHT states specified by $\{M,J\}$. Their BHT follows from the Bekenstein-Hawking entropy formula Eq. (\ref{10}), which yields the entropy $S=S(M,J)$ on evaluating the area $A=A(M,J)$ of the event horizon $\mathcal{H}$ with GR.

\par
The area calculation starts with the Kerr line element in Boyer-Lindquist coordinates $(t,r,\theta,\phi)$ in natural units $G=c=1$, with $G$ the constant of gravitation and $c$ the speed of light \cite{Carroll2019}:

\begin{equation}\label{E: Kerr Metric}
    ds^2 = -\bigg(1 - \frac{2 M r}{\rho^2}\bigg)dt^2 -  \frac{4 M a\,r \sin^2{\theta}}{\rho^2} dt\; d\phi + \frac{\rho^2}{\Delta} dr^2 + \rho^2 d\theta^2 + \frac{\sin^2{\theta}}{\rho^2} \bigg((r^2 + a^2)^2 - a^2 \Delta \sin^2{\theta}\bigg)d\phi^2.
\end{equation}

\noindent Here, $\rho^2 = r^2 + a^2\cos^2\theta$, the discriminant $\Delta = r^2 - 2Mr + a^2$, and $a = J/M$. The outer event horizon radius $r_+$ is determined by solving $\Delta = 0$ for the larger of the two real valued solutions. For the Kerr black hole, this gives $r_+ = M + \sqrt{M^2 - a^2}$. For $r_+$ to be real, we clearly require $M^2>a^2$. The case $M=a$ marks the extremal limit. 

\par 
To find the area of the static event horizon, begin by setting $dt = dr = 0$ and $r = r_+$. The resulting line element is that of the Kerr black hole event horizon,

\begin{equation}
    ds_+^2 = \bigg(r_+^2 + a^2 + \frac{2Mr_+a^2 \sin^2\theta}{\rho_+^2} \bigg) \sin^2\theta d\phi^2 + \rho_+^2 d\theta^2,
\end{equation}

\noindent where the $+$ subscript denotes a dependence on $r_+$. The determinant of this metric is $\det{g_+} = (2Mr_+\sin \theta)^2$. Finally, we may compute the area of the event horizon,

\begin{equation}
    A = \int_{\mathcal{H}} d\Omega = \int_{\mathcal{H}}\sqrt{\det g_+}d\theta d\phi = \int_{\mathcal{H}} 2Mr_+ \sin \theta d\theta d\phi = 8\pi Mr_+.
\end{equation}

\noindent Eq. (\ref{10}) now gives $S(M,J)$. We used the scaling of {\AA}man et al. \cite{Aaman2003}, who set $L_p=1$, $k_B=1$, and employed dimensionless units for $\{M,J\}$. These authors also inserted an extra dividing scaling factor of $\pi$ to get the entropy of the Kerr black hole:

\begin{equation} \label{E: Kerr Entropy}
    S(M, J) = 2 M^2\left(1+\sqrt{1-\frac{J^2}{M^4}}\right).
\end{equation}

\noindent The temperature follows from Eq. (\ref{E: Temperature definition}):

\begin{equation}
    \frac{1}{T}=\frac{4 M^3}{\sqrt{M^4-J^2}}+4 M.\label{KerrTemp}
\end{equation}

\noindent There is a maximum temperature for any given $M$: $T=1/8M$. This temperature corresponds to $J=0$ for Kerr, as seen in Eq. (\ref{KerrTemp}). Maximum $M$ dependent temperatures are evident in a number of our contour plots below.

\par
The thermodynamic metric is constructed from $S(M,J)$ with Eq. (\ref{E: thermo metric}). With Eq. (\ref{RicciScalar}) the scalar thermodynamic curvature $R$ is found to be \cite{Aaman2003}:

\begin{equation}
    R=\frac{1}{4 M^2}\frac{2-\sqrt{1-\frac{J^2}{M^4}}}{\sqrt{1-\frac{J^2}{M^4}}}.
    \label{596032}
\end{equation}

\noindent Clearly, $R$ is positive for of all the physical states. $R$ in Eq. (\ref{596032}) is the negative of that in \cite{Aaman2003} because of opposite $R$ sign conventions. 

\par
The only singularity in $R$ occurs at the extremal limit where $J\to M^2$, and $T\to 0$. The limiting form is:

\begin{equation} \label{E: Kerr Scalar Result}
    R\to\frac{1}{8 M^3 T},
\end{equation}

\noindent first found by Ruppeiner \cite{Ruppeiner2008} using a slightly different scaling. In sign and $T$ dependence, this limiting form matches Eqs. (\ref{limitingline1}) and (\ref{limitingline2}) for the 2D and the 3D ideal Fermi gasses, respectively. But the coefficients of $1/T$ for Kerr and ideal Fermi differ from one another in their mass dependences, possibly pointing to the need for a better ideal Fermi model to make a full correspondence between OT and Kerr BHT. This is a project for the future.

\par
Clearly for the proportionality $R\propto +1/T$ to hold for Kerr requires constant $M$, from Eq. (\ref{E: Kerr Scalar Result}). Guided by this, we examine $R$ only along lines of constant $M$ for all our models, a procedure which produces excellent results. We note that Kerr is thermodynamically unstable everywhere \cite{Ruppeiner2007}, including near the extremal limit, and this may diminish its interest. Kerr has a simple closed form solution for $R$, but other cases considered here have $R$ consist of possibly hundreds or even thousands of terms. Therefore, our main calculational effort must be numerical.

\par
For Kerr, Figure \ref{F: 3D Plots}(a) shows a contour plot of $R$ in the $\{M,T\}$ plane. The key point to notice is that $R$ increases as $T$ decreases, as anticipated from the limiting expression Eq. (\ref{E: Kerr Scalar Result}). The analytic expression for $R$ indicates that $R$ for Kerr diverges only as $T\to 0$, so Kerr is devoid of the second-order phase transitions seen in other BHT models away from the extremal line. As Eq. (\ref{KerrTemp}) for $T$ shows, all positive values of $M$ allow $T\to 0$ (with $J\to M^2$).

\par
Another look at the temperature dependence of the scalar curvature for Kerr is found in Figure \ref{F: Fitted Plots}(a), which shows $R$ as a function of $T$, with $M$ fixed at various values. On a log-log scale, an asymptotic $R\propto 1/T$ relation presents as a straight line with slope $-1$. This is indeed the case in the figure, for all $M$. For small $T$, the values of $R$ agree with those in Eq. (\ref{E: Kerr Scalar Result}). Fit values of our coefficients are shown in Table \ref{fit coefficient table}.

\par
Let us make one additional point. In \cite{Ruppeiner2008} it was found that in the extremal limit the product of $R$ and the heat capacity at constant $J$ and $Q$ goes to unity for $\{J,Q\}$, $\{M,J\}$, and $\{M,Q\}$ fluctuations. Exactly the analogous behaviour was found for the 2D ideal Fermi gas. This points to an additional connection between BHT and the 2D ideal Fermi gas. However, we make no attempt to generalize this find here because of the difficulty of evaluating heat capacities in BHT and because of uncertainties about the appropriate heat capacities to use in more complex models. In this survey, we confine ourselves to analyzing just the extremal invariant $R$, whose appropriateness is never in doubt.

\subsection{Kerr-Newman $\{M, \, J\}$ $(Q=0.4)$} \label{SS: KNMJ}

The three parameters $\{M, J, Q\}$ characterizing Kerr-Newman (KN) black holes \cite{Carroll2019} provide a rich avenue of exploration into the BHT thermodynamic geometry. The entropy's dependence on these three parameters gives rise to seven different thermodynamic geometries based on which (if any) of the three parameters are held fixed \cite{Ruppeiner2007, Sahay2017b, Ruppeiner2008}. But we restrict ourselves in this paper to exactly two of the three $\{M, J, Q\}$ fluctuating, $\{M,J\}$ and $\{M,Q\}$, both with fluctuating $M$. We omit $\{J,Q\}$ fluctuations since they do not have fluctuating $M$. For the case of all three parameters $\{M, J, Q\}$ fluctuating see \cite{Ruppeiner2008, Mirza2007}.

\par
In the scaling of {\AA}man et al. \cite{Aaman2003} the Kerr-Newman BHT has entropy function:

\begin{equation}
   S(M, J, Q) = 2 M^2 - Q^2 + 2M^2\sqrt{1-\frac{J^2}{M^4}-\frac{Q^2}{M^2}}.
   \label{KNentropy}
\end{equation}

\noindent The temperature follows from Eq. (\ref{E: Temperature definition}):

\begin{equation}
    \frac{1}{T}=\frac{2(K^2+2K+L^2)M}{K},
    \label{KNTemp}
\end{equation}

\noindent where the variables

\begin{equation}
    \{\alpha,\beta\}=\left\{\frac{J^2}{M^4},\frac{Q^2}{M^2}\right\},
\end{equation}

\noindent and

\begin{equation}
    \{K,L\}=\{\sqrt{1-\alpha-\beta},\sqrt{1+\alpha}\}.
\end{equation}\

\par
In this subsection, we consider $\{M,J\}$ fluctuating at fixed $Q>0$ ($Q=0$ is simply Kerr). The entropy and the temperature are given by Eqs. (\ref{KNentropy}) and (\ref{KNTemp}), respectively. Eq. (\ref{E: thermo metric}) yields the thermodynamic metric elements, and Eq. (\ref{RicciScalar}) yields the thermodynamic curvature:

\begin{equation} R = \frac{\begin{array}{lll}
(K^7+3 K^6+2 K^5 L^2+6 K^4 L^2-5 K^4+K^3 L^4+9 K^3 L^2-\\\qquad 9 K^3+3 K^2 L^4+4 K^2 L^2-8K^2+9 K L^4-\\\qquad \qquad 21 K L^2+12 K+9 L^4-24 L^2+16)\end{array}}{2 K M^2 \left(2 K^3+3 K^2+2 K L^2-2 K+3L^2-4\right)^2}.\end{equation}

\noindent It is straightforward to show that the limiting $R$ at small $T$ is:

\begin{equation}R\to\frac{1}{4 L^2 M^3 T},\label{10986}\end{equation}

\noindent so again $R\propto +1/T$ in the extremal limit along lines of constant $M$.

\par
\begin{figure} [H]
    \begin{tabular}{*{3}{@{}c}@{}}
        \includegraphics[width = 0.33\linewidth]{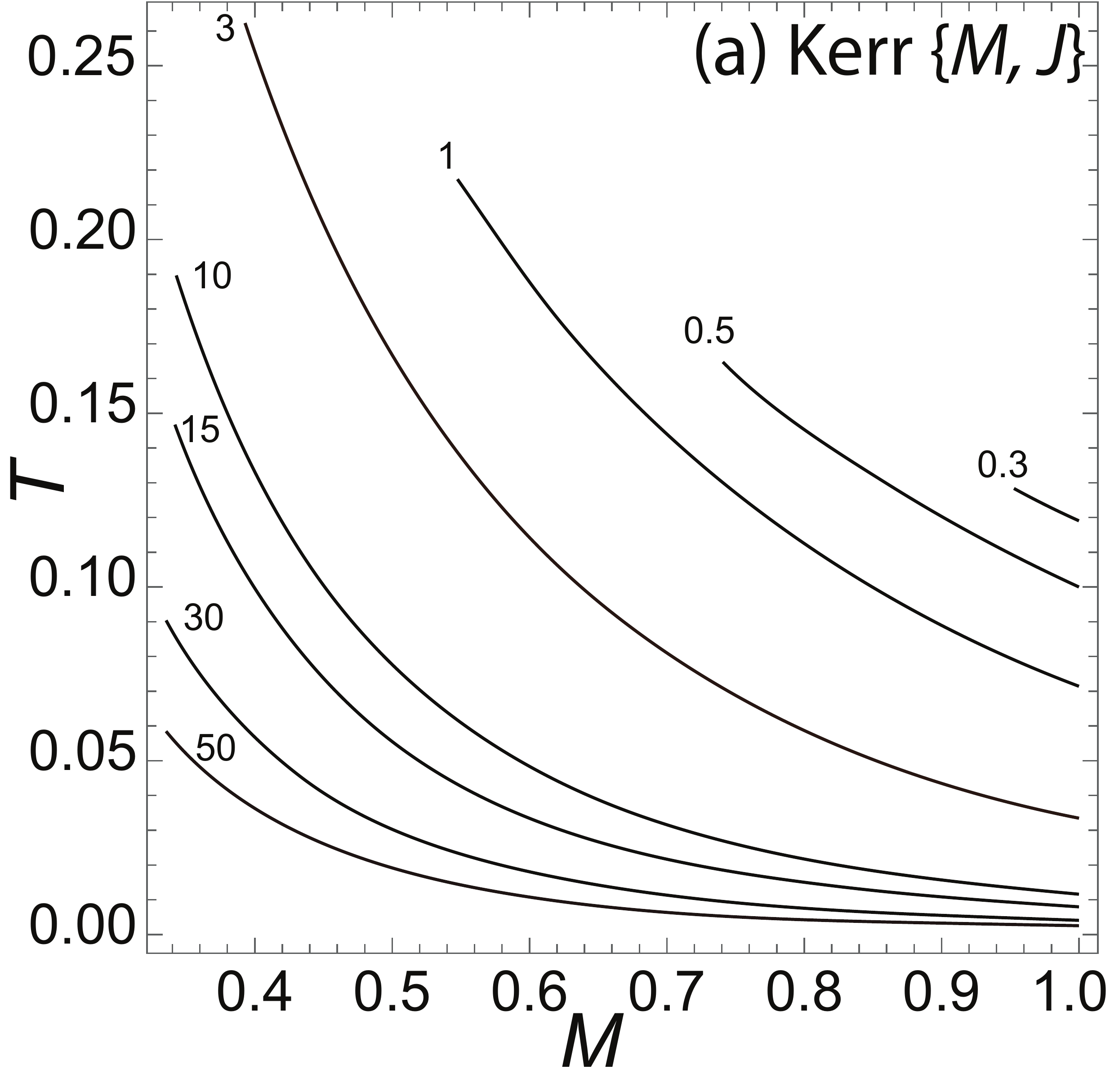} &
        \includegraphics[width = 0.33\linewidth]{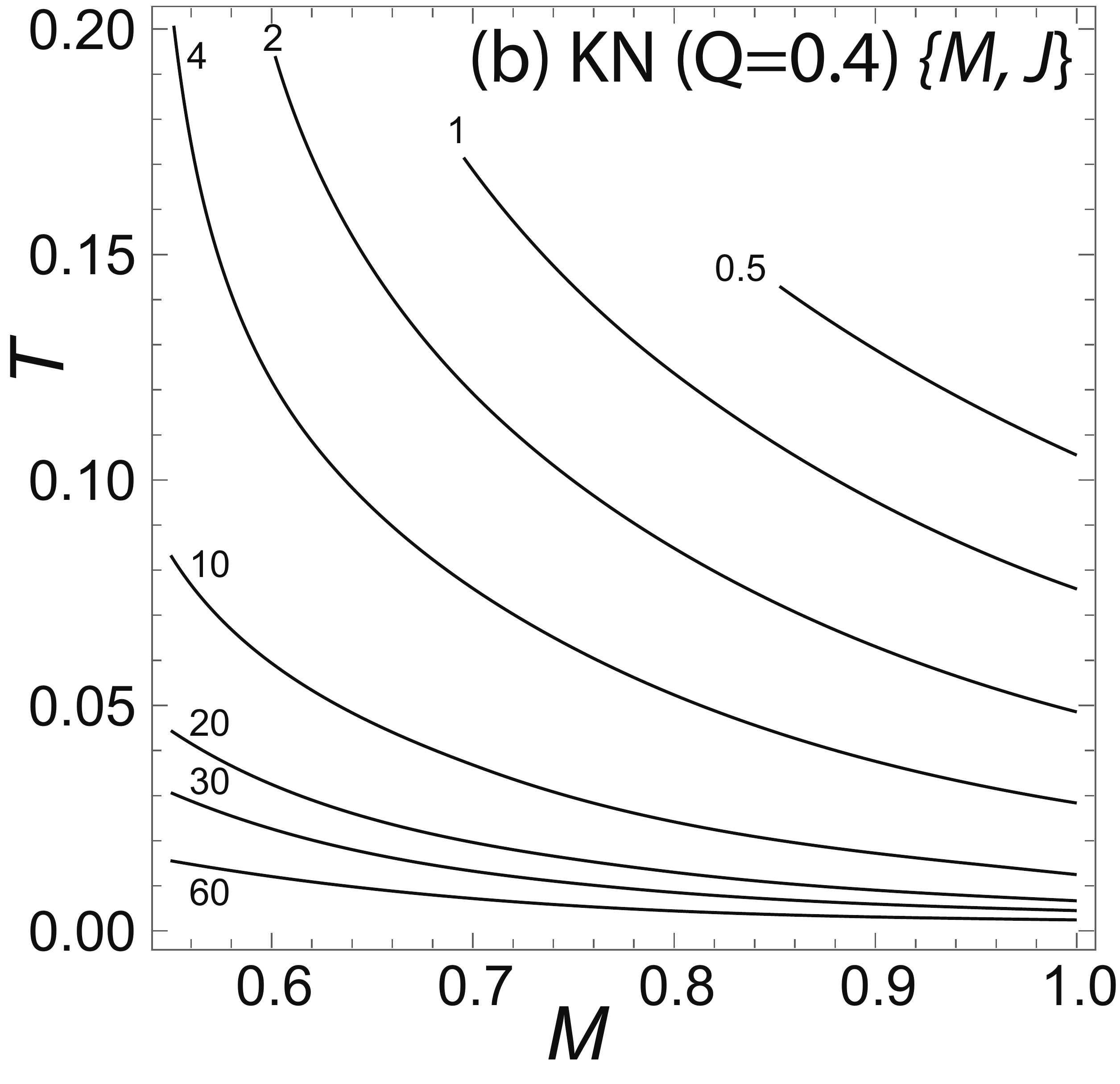} &
        \includegraphics[width = 0.33\linewidth]{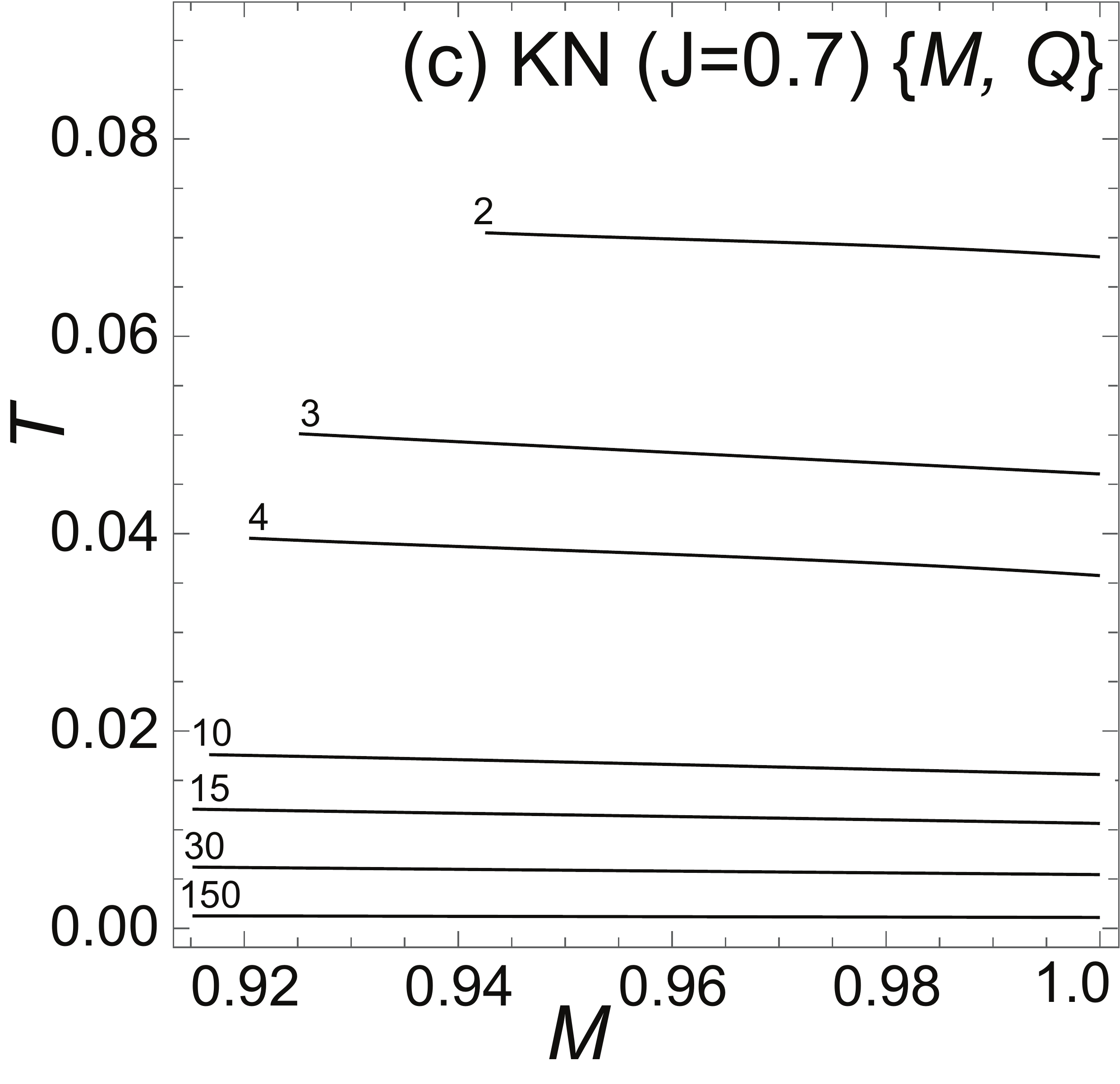}
        \\
        \includegraphics[width = 0.33\linewidth]{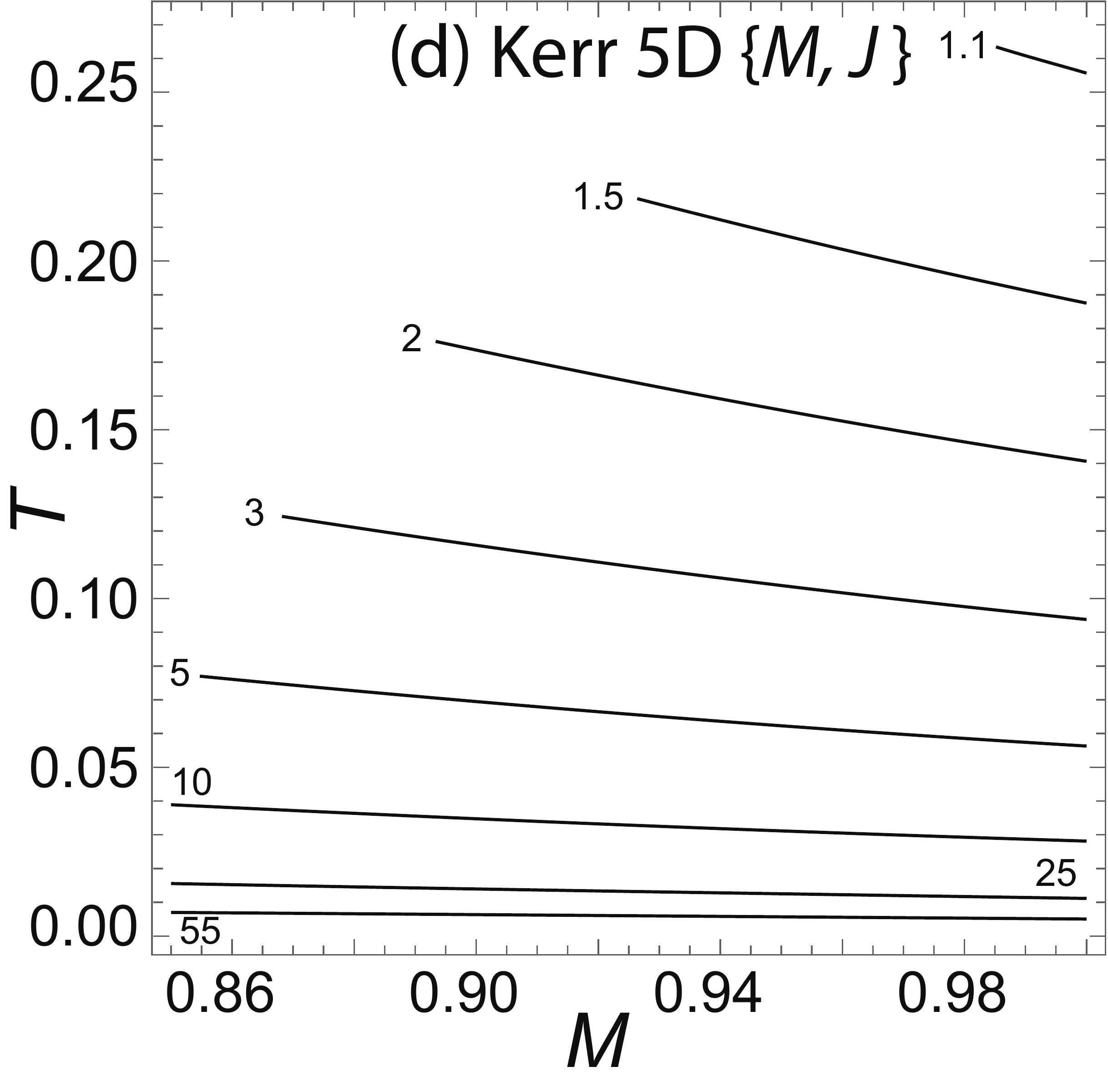} &
        \includegraphics[width = 0.33\linewidth]{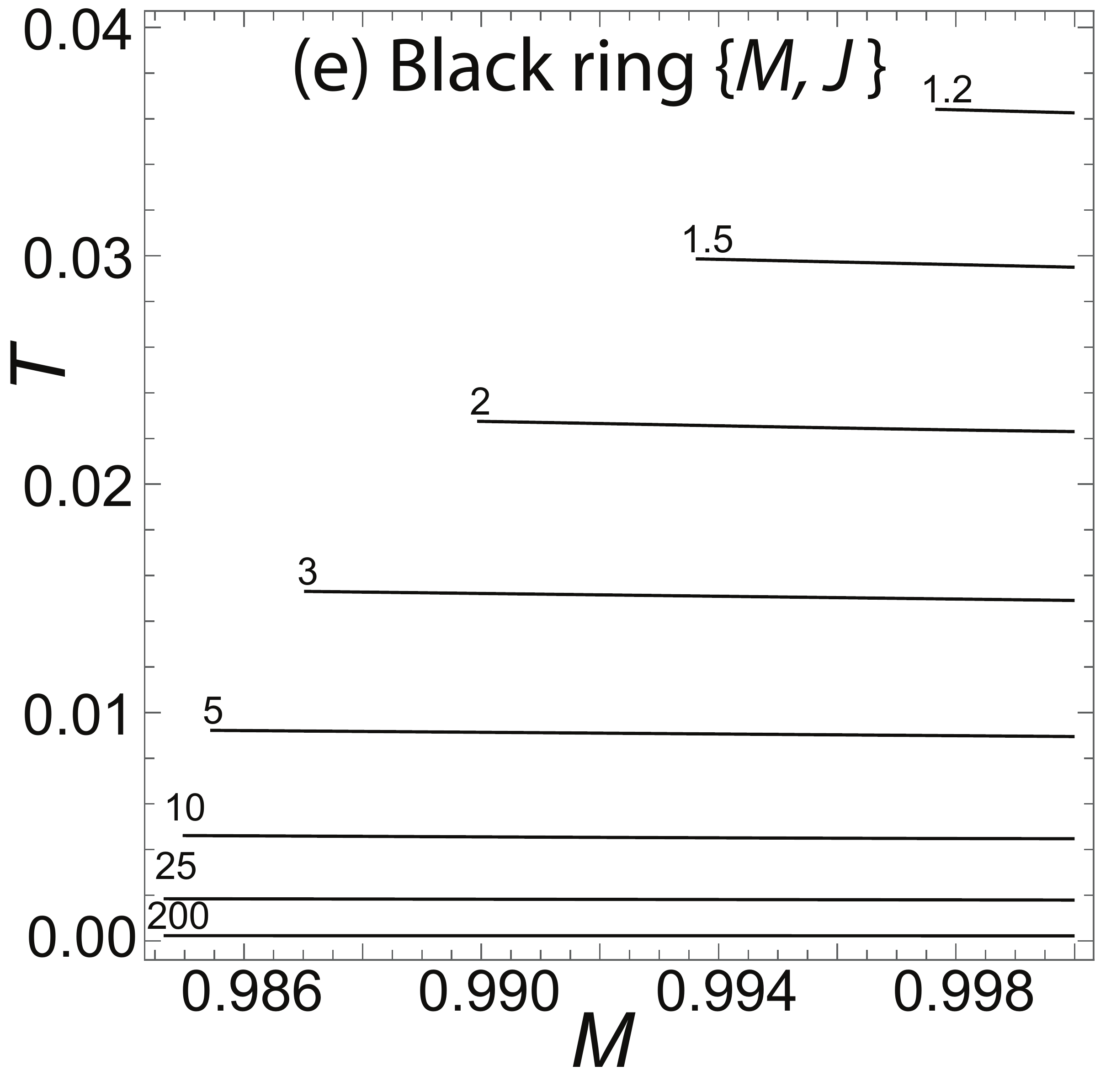} &
        \includegraphics[width = 0.33\linewidth]{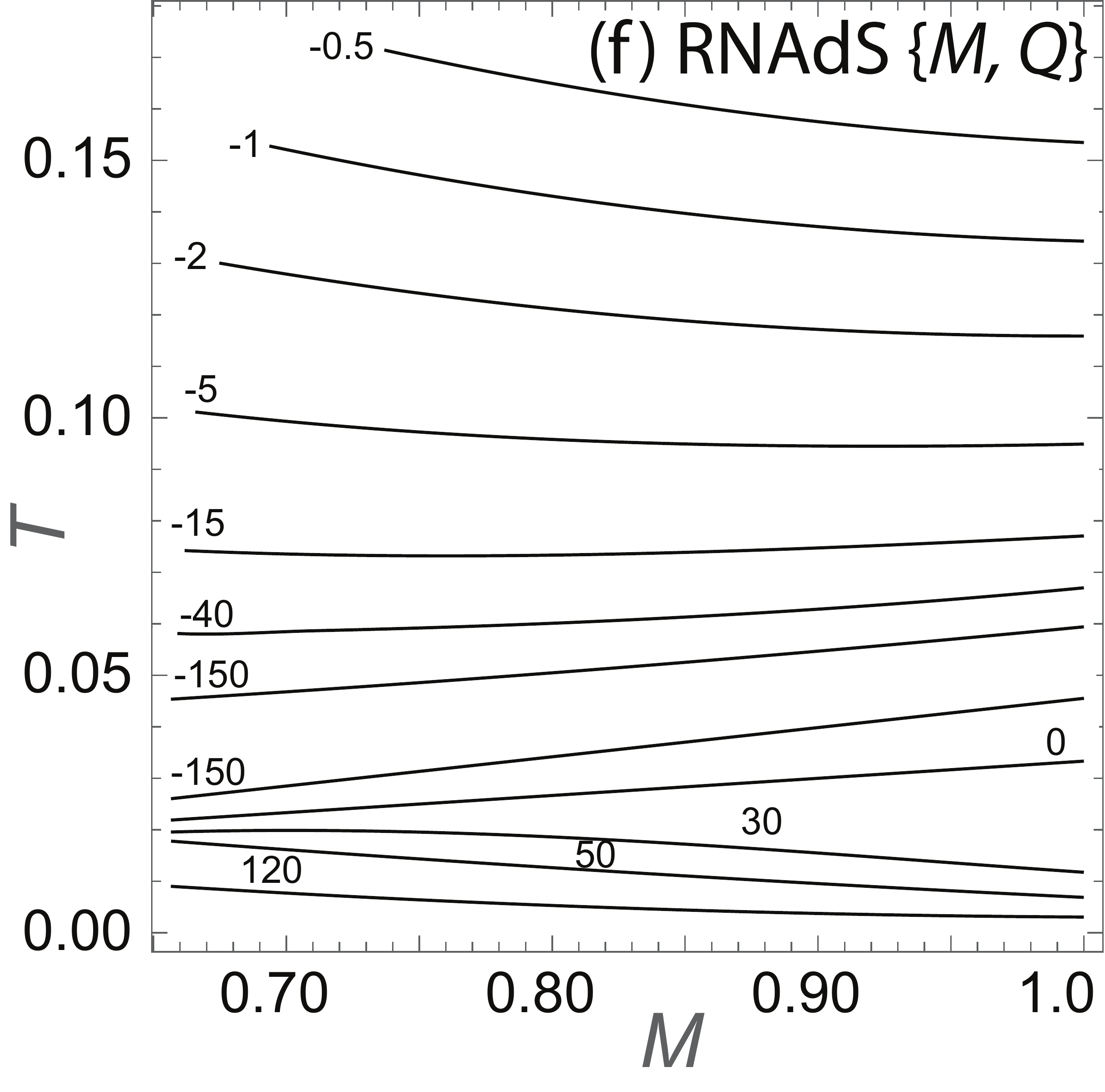}
        \\
        \includegraphics[width = 0.33\linewidth]{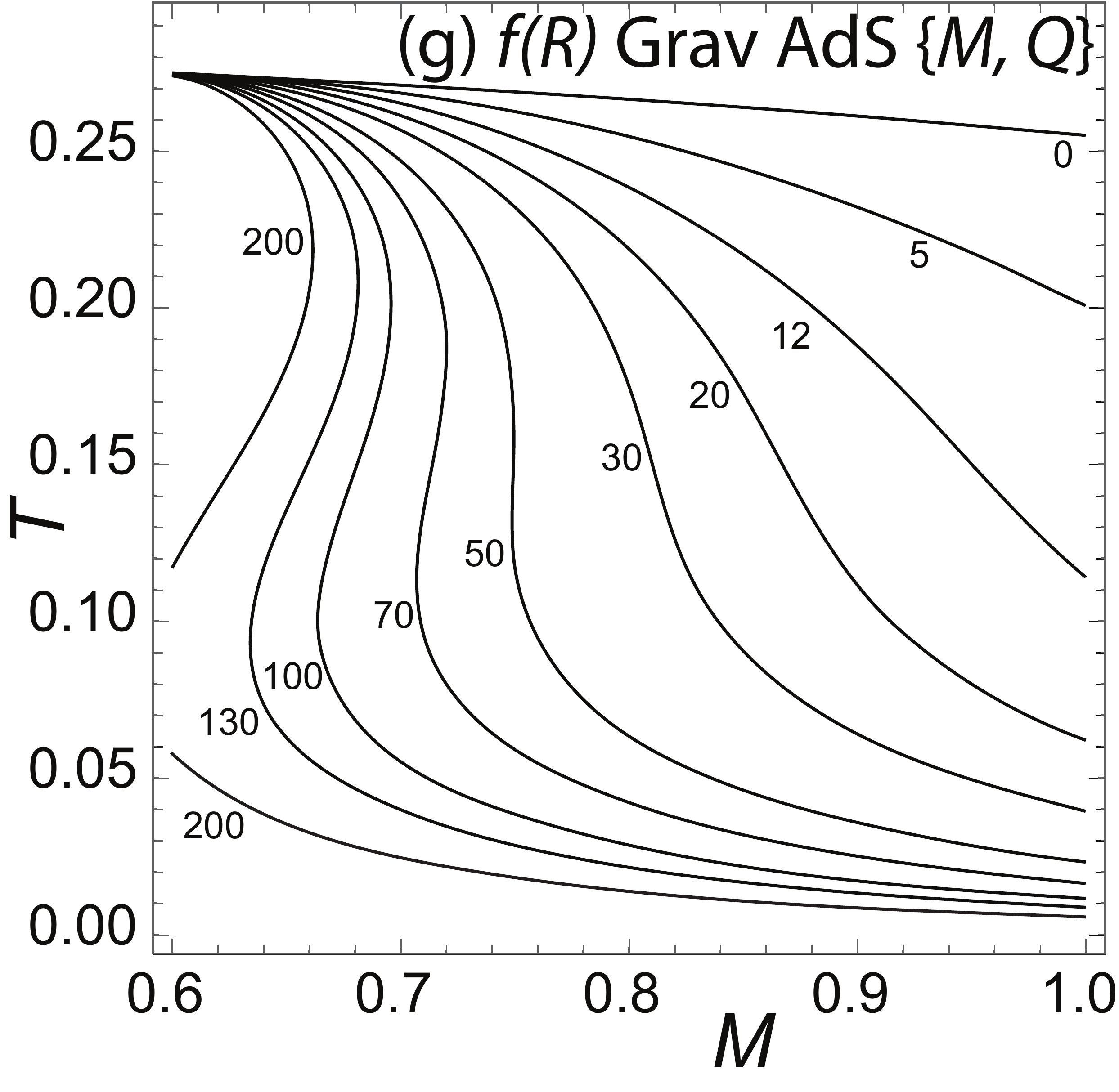} &   \includegraphics[width = 0.33\linewidth]{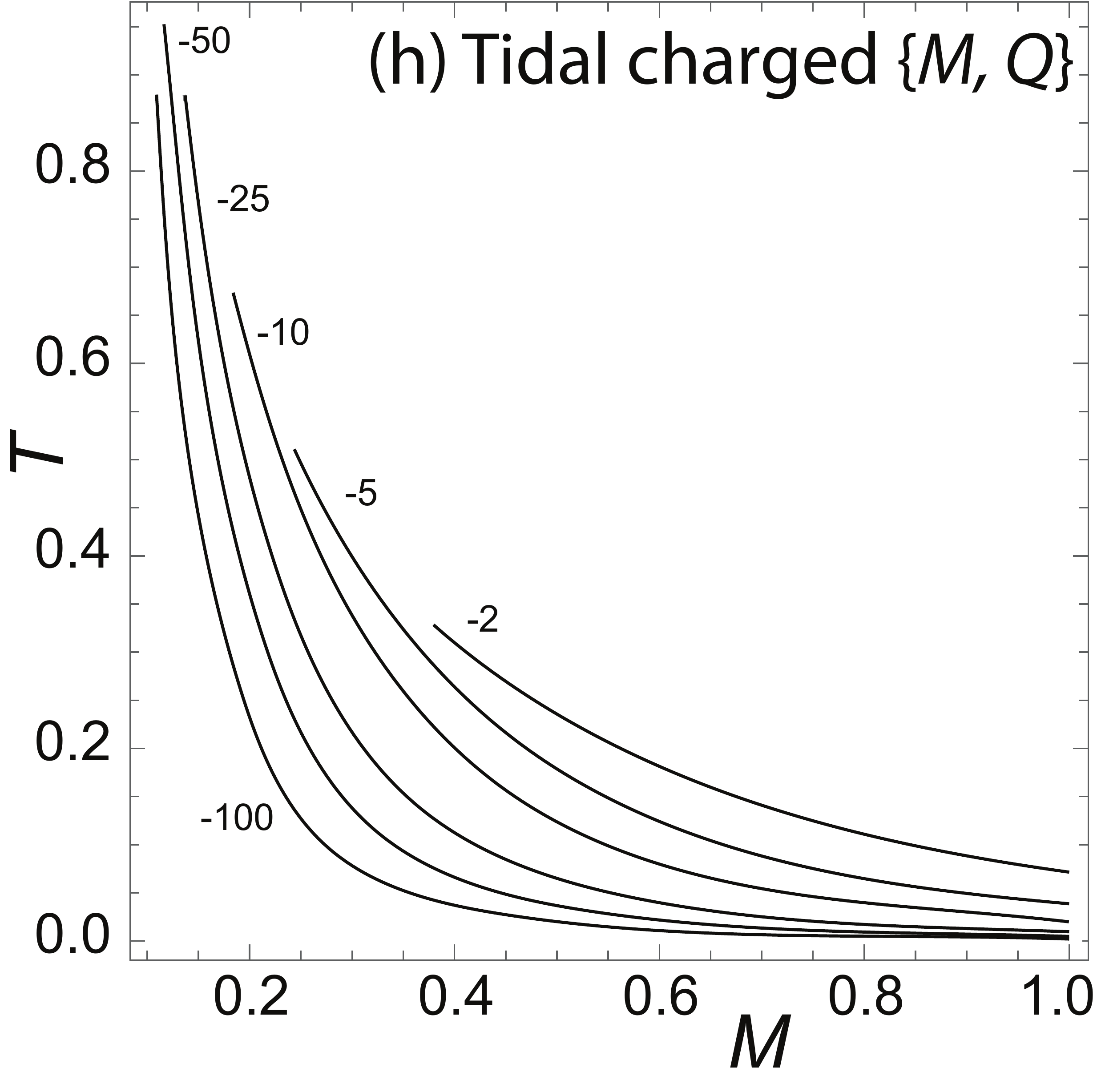} &
        \includegraphics[width = 0.33\linewidth]{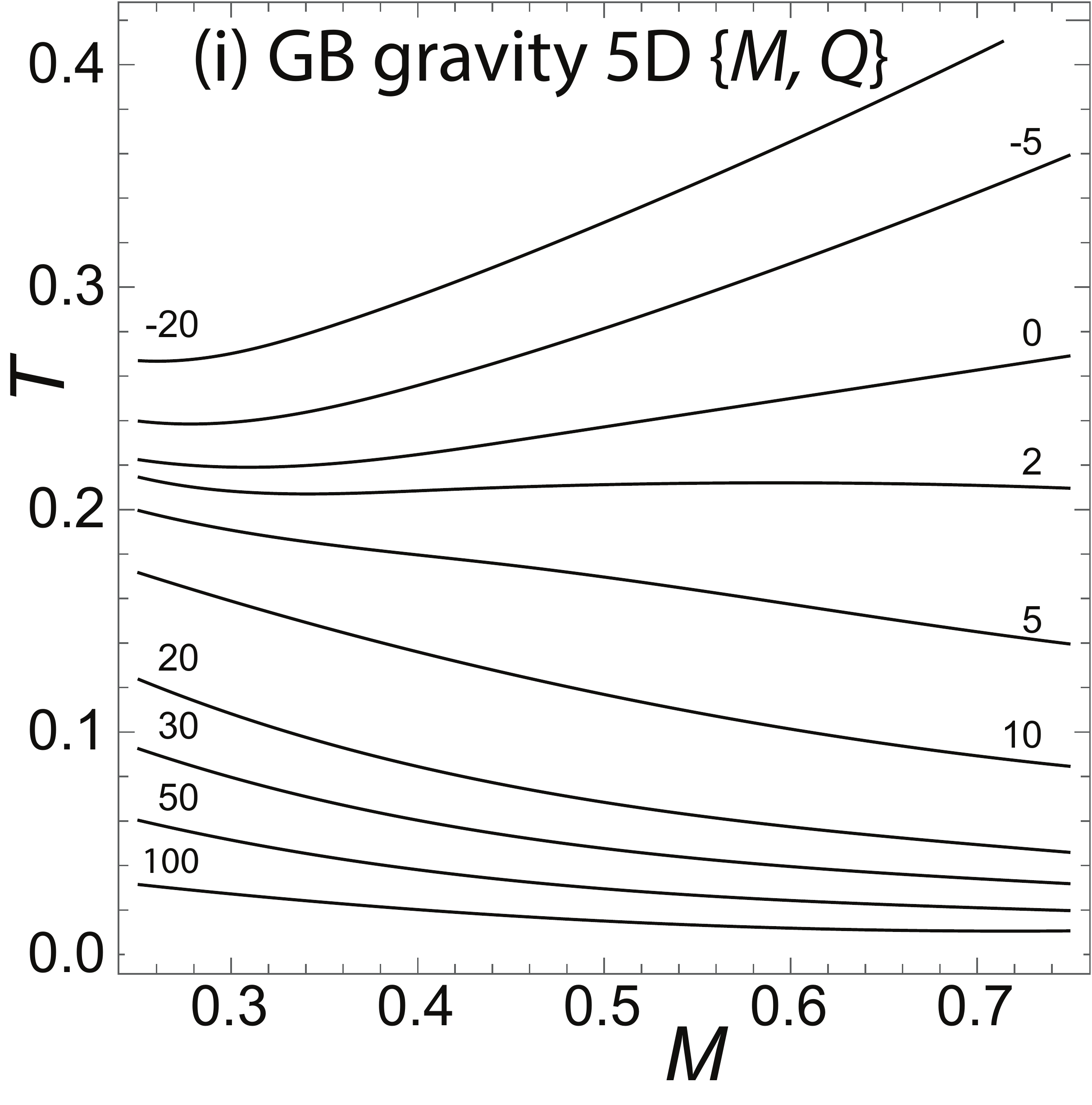}
        \\
        \includegraphics[width = 0.33\linewidth]{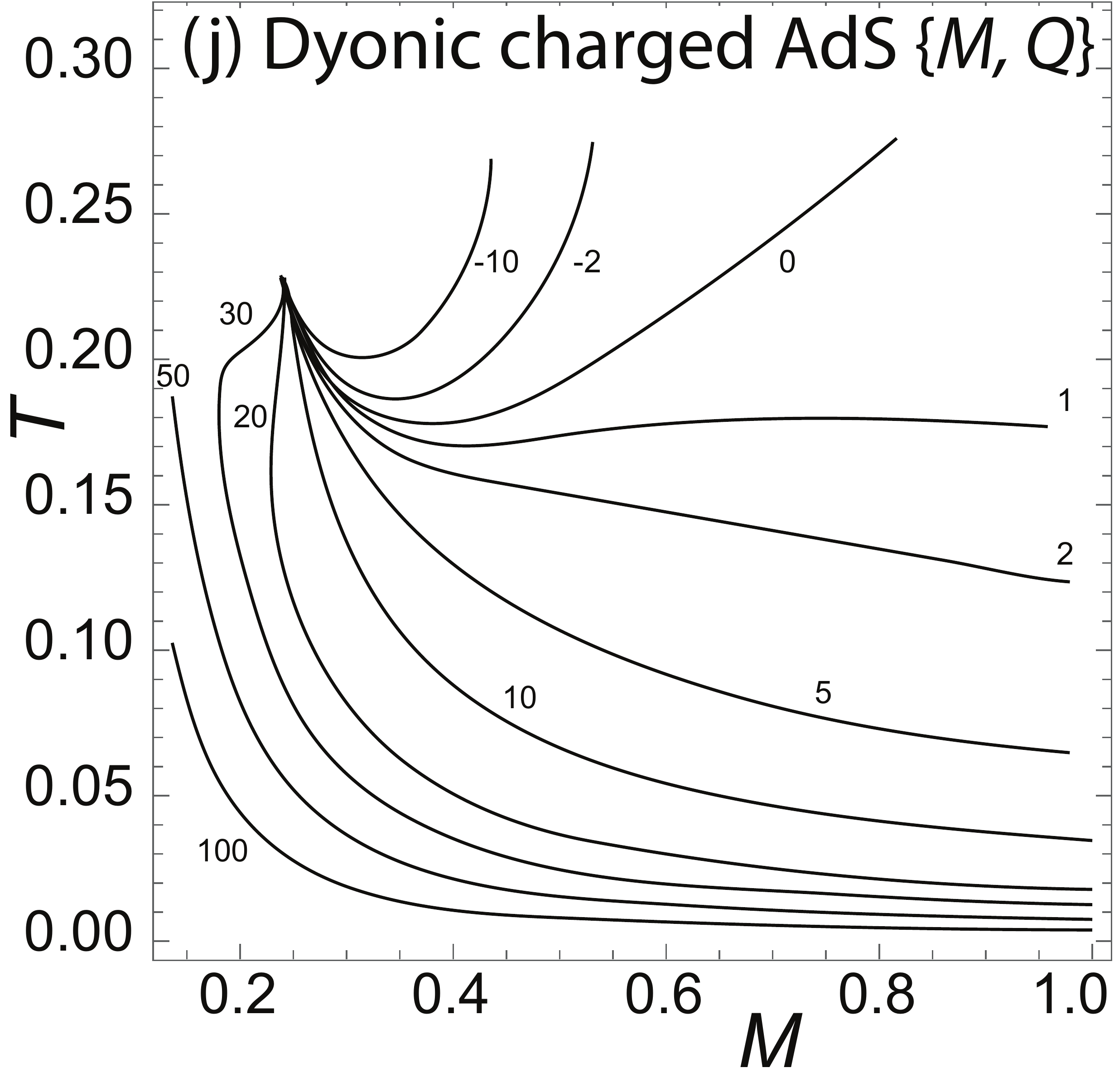} &
        \includegraphics[width = 0.33\linewidth]{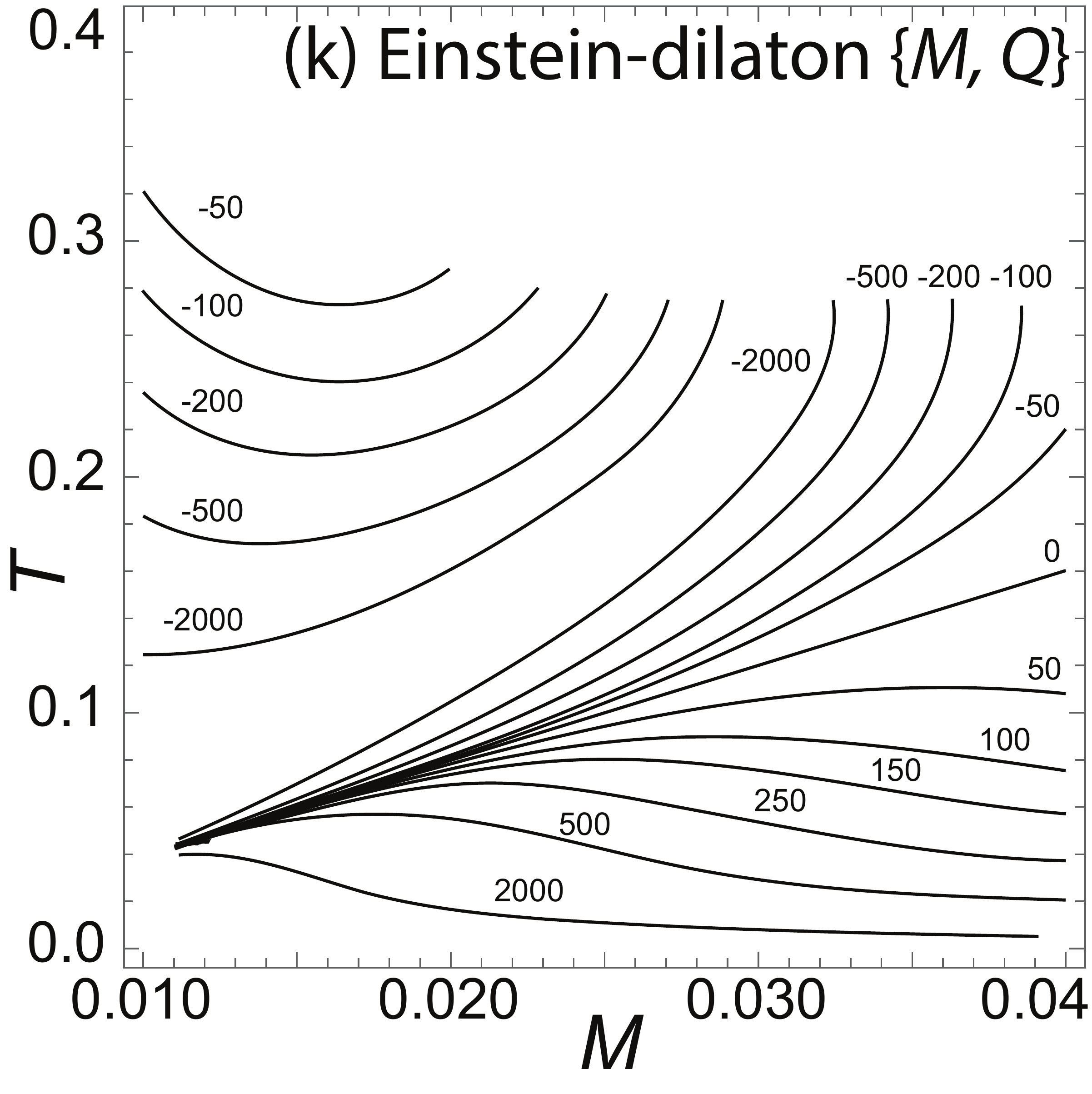} & \includegraphics[width = 0.33\linewidth]{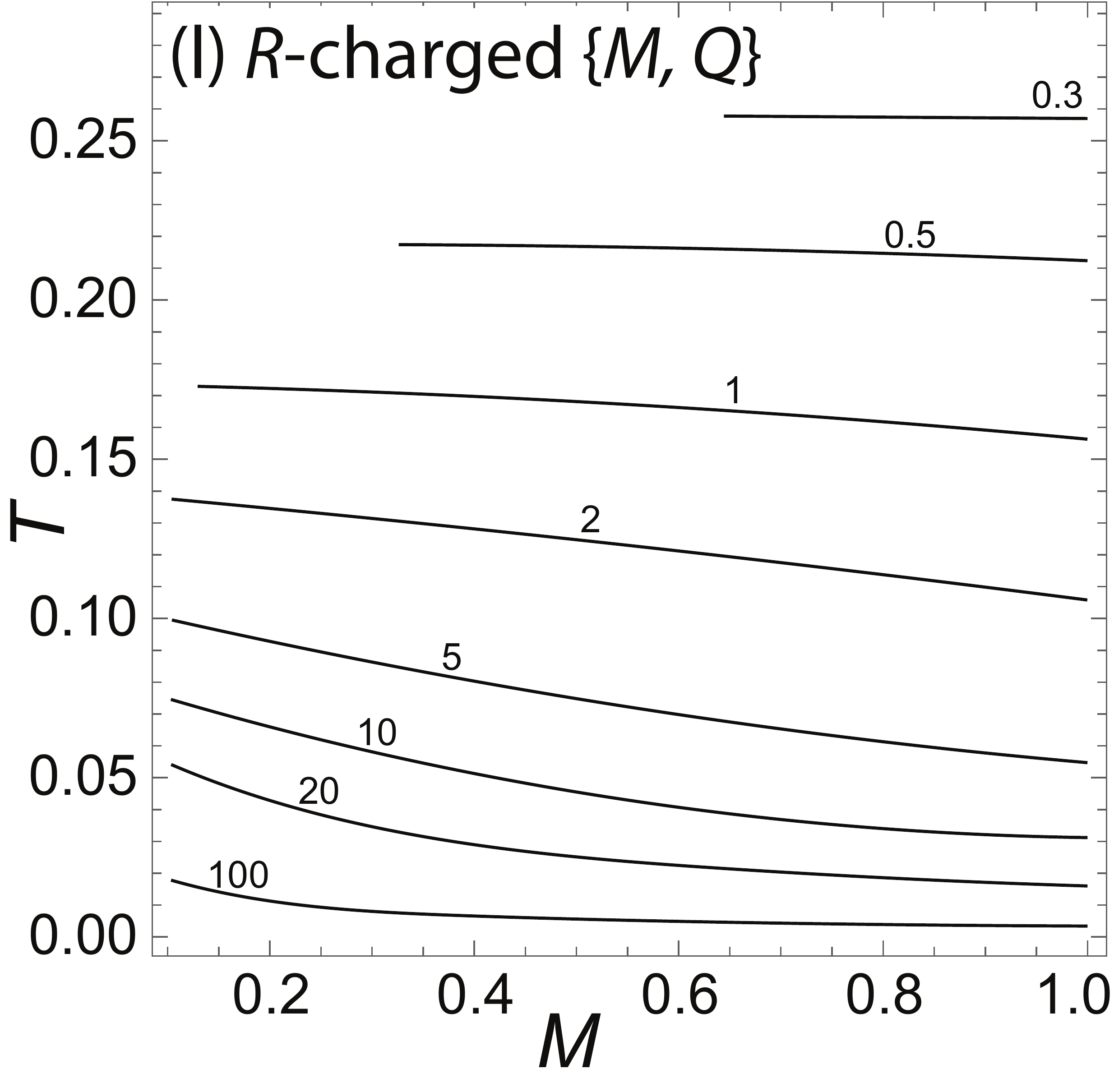}\\
    \end{tabular}
    \caption{Contour plots for $R$ for the twelve BHT models considered.}
    \label{F: 3D Plots}
\end{figure}

Figure \ref{F: 3D Plots}(b) shows a contour plot of $R$ in the $\{M,T\}$ plane, for fixed $\{Q,\Lambda\}=\{0.4,0\}$. Generally, $R$ is seen to increase as $T$ decreases to zero. For $Q=0.4$, $R$ is uniformly positive, and has no divergences other than in the extremal limit \cite{Ruppeiner2008}. The BHT has uniformly unstable thermodynamics for $Q=0.4$ \cite{Ruppeiner2007}, and this may diminish the interest in this model. The contour lines in Fig. \ref{F: 3D Plots}(b) each terminate at an $M$ dependent upper limiting value for $T$.

\begin{figure}[H]
\begin{tabular}{*{3}{@{}c}@{}}
    \includegraphics[width = 0.33\linewidth]{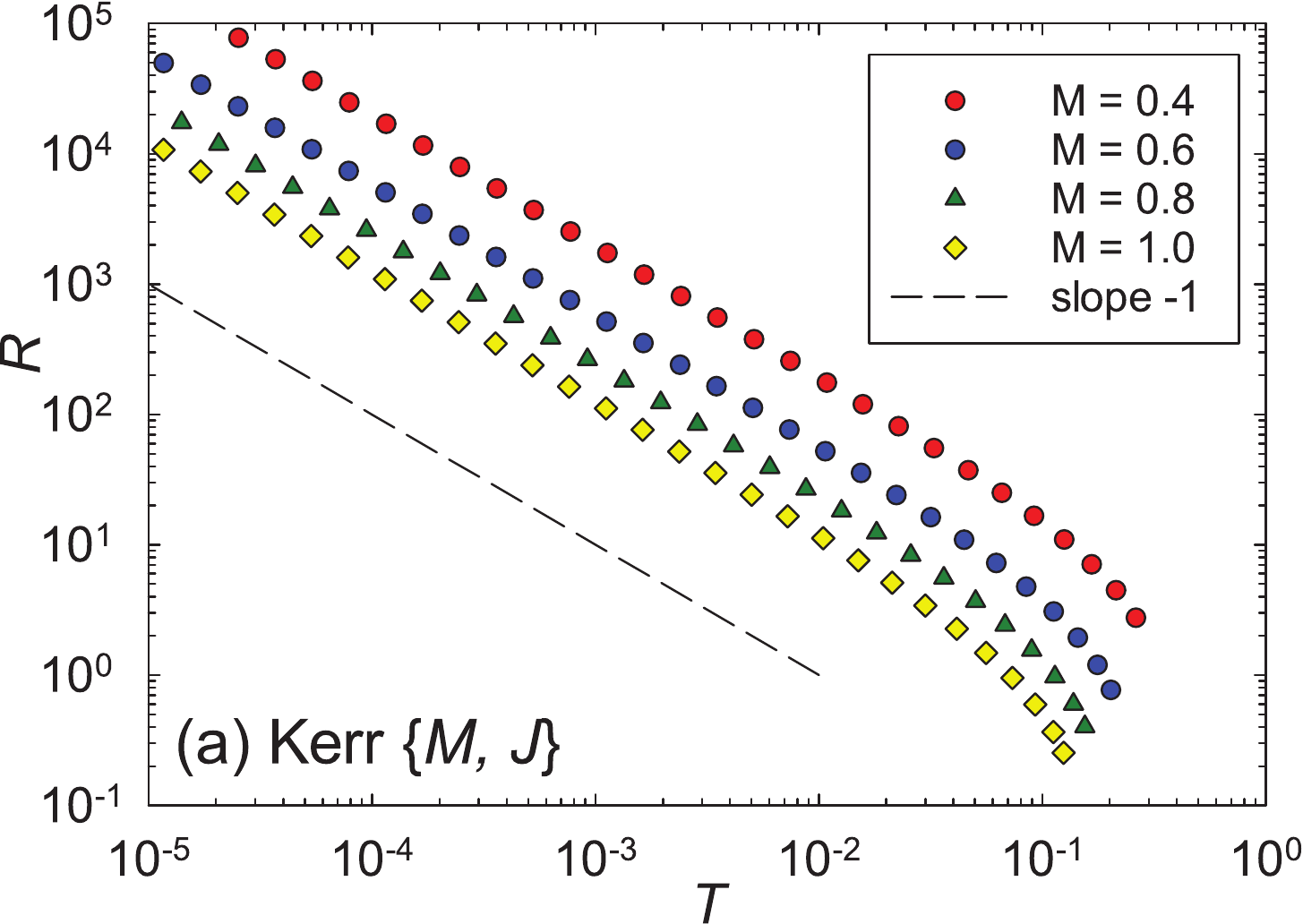} &
    \includegraphics[width = 0.33\linewidth]{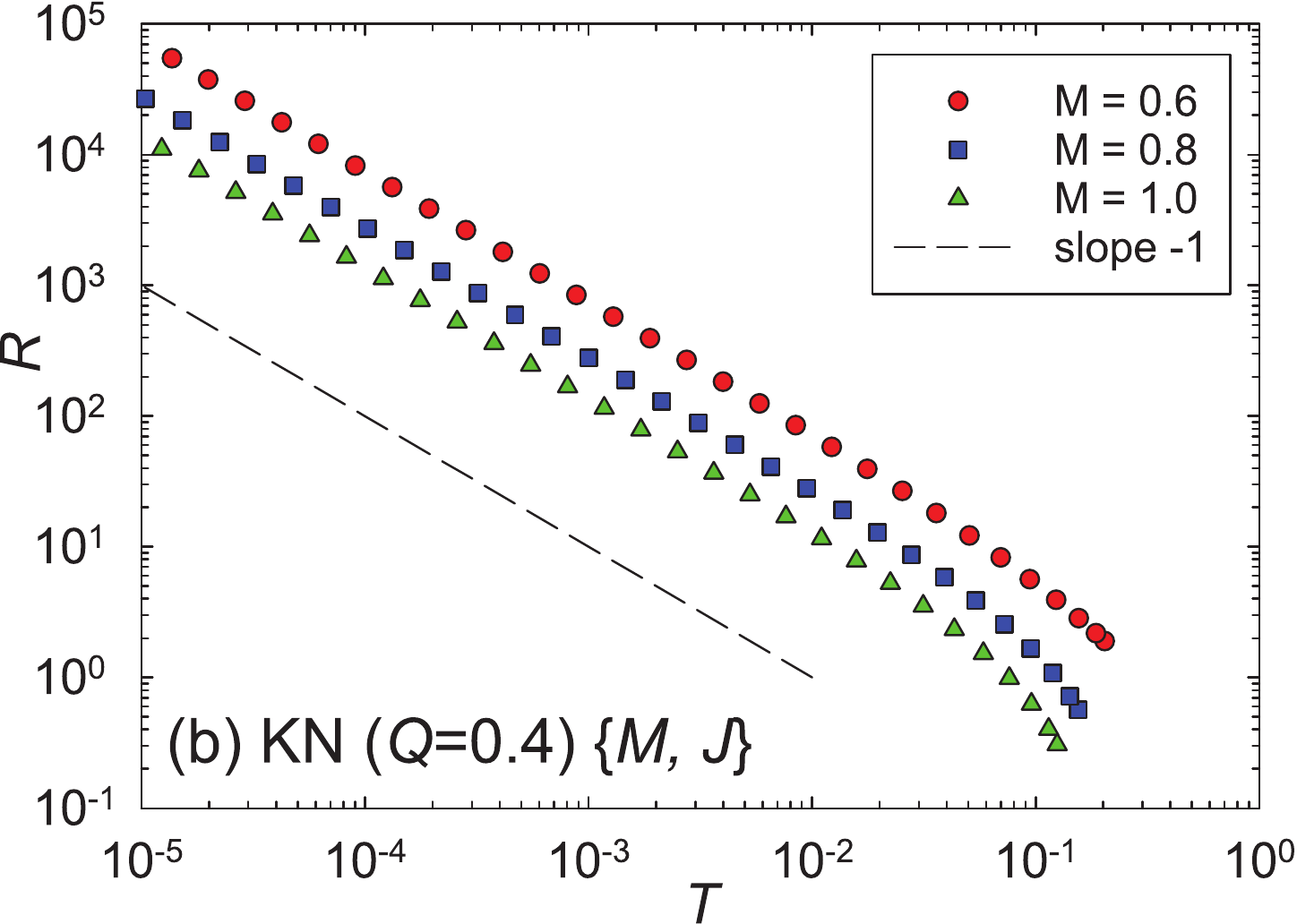} &
    \includegraphics[width = 0.33\linewidth]{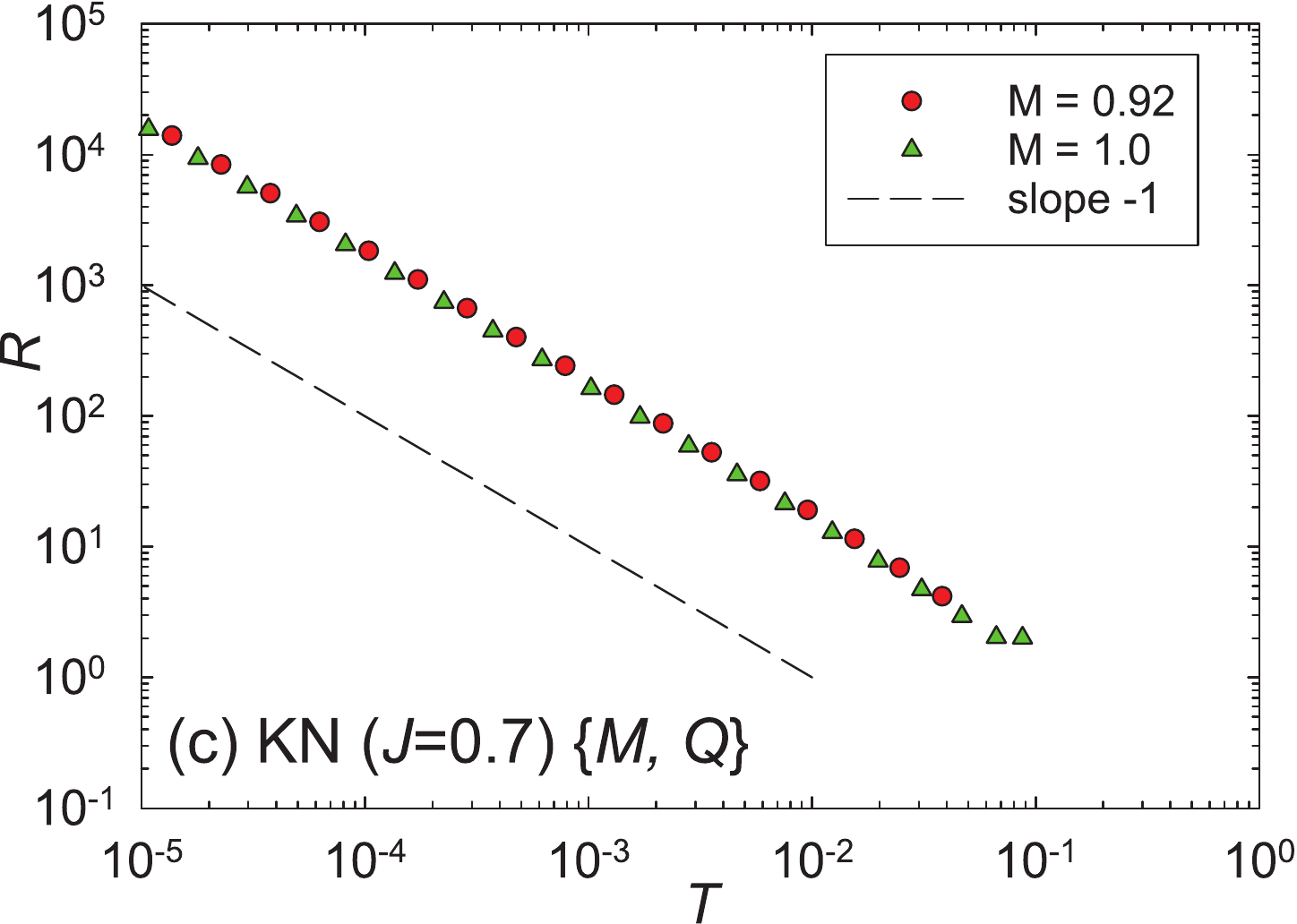}
    \\
    \includegraphics[width = 0.33\linewidth]{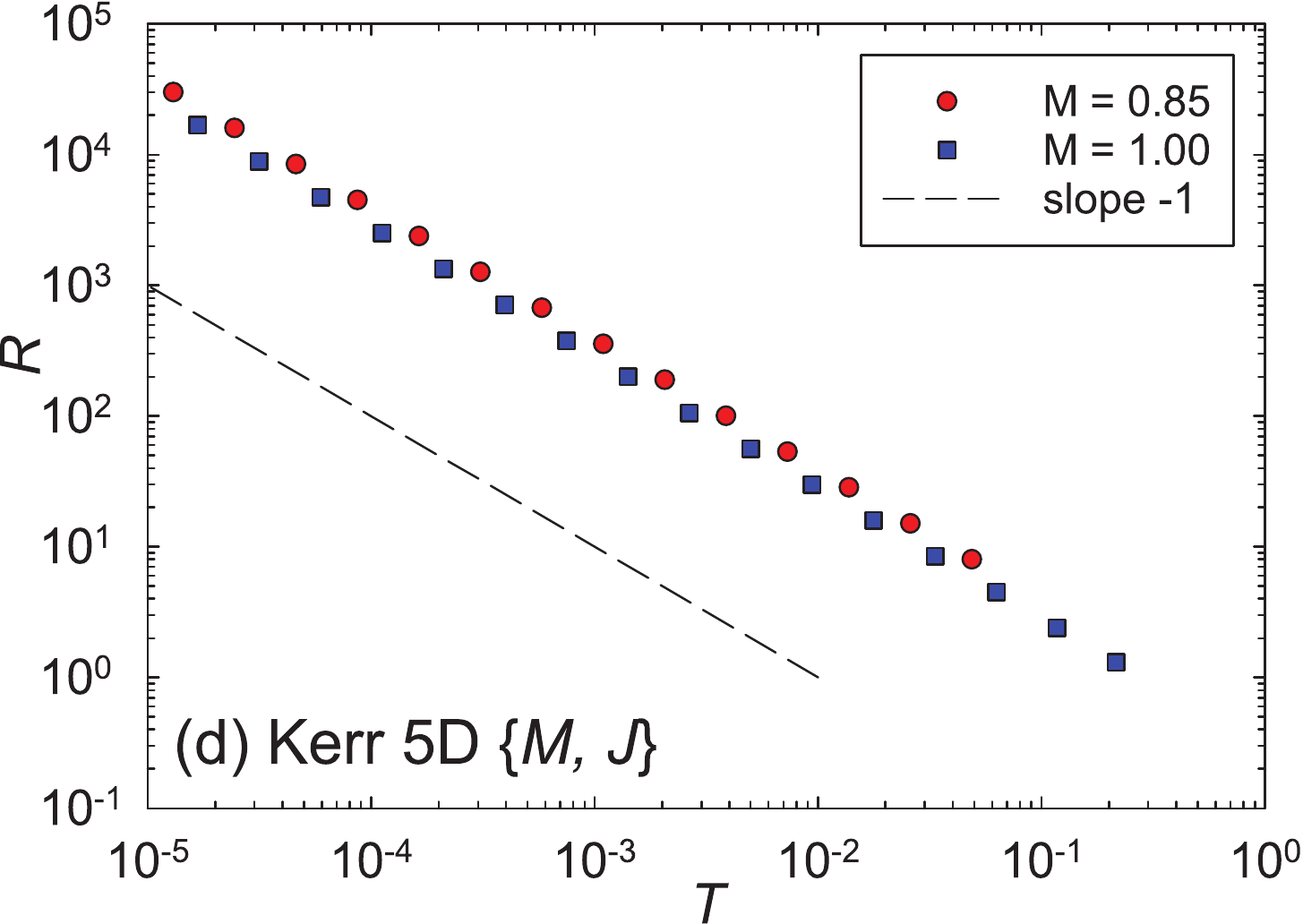} &
    \includegraphics[width = 0.33\linewidth]{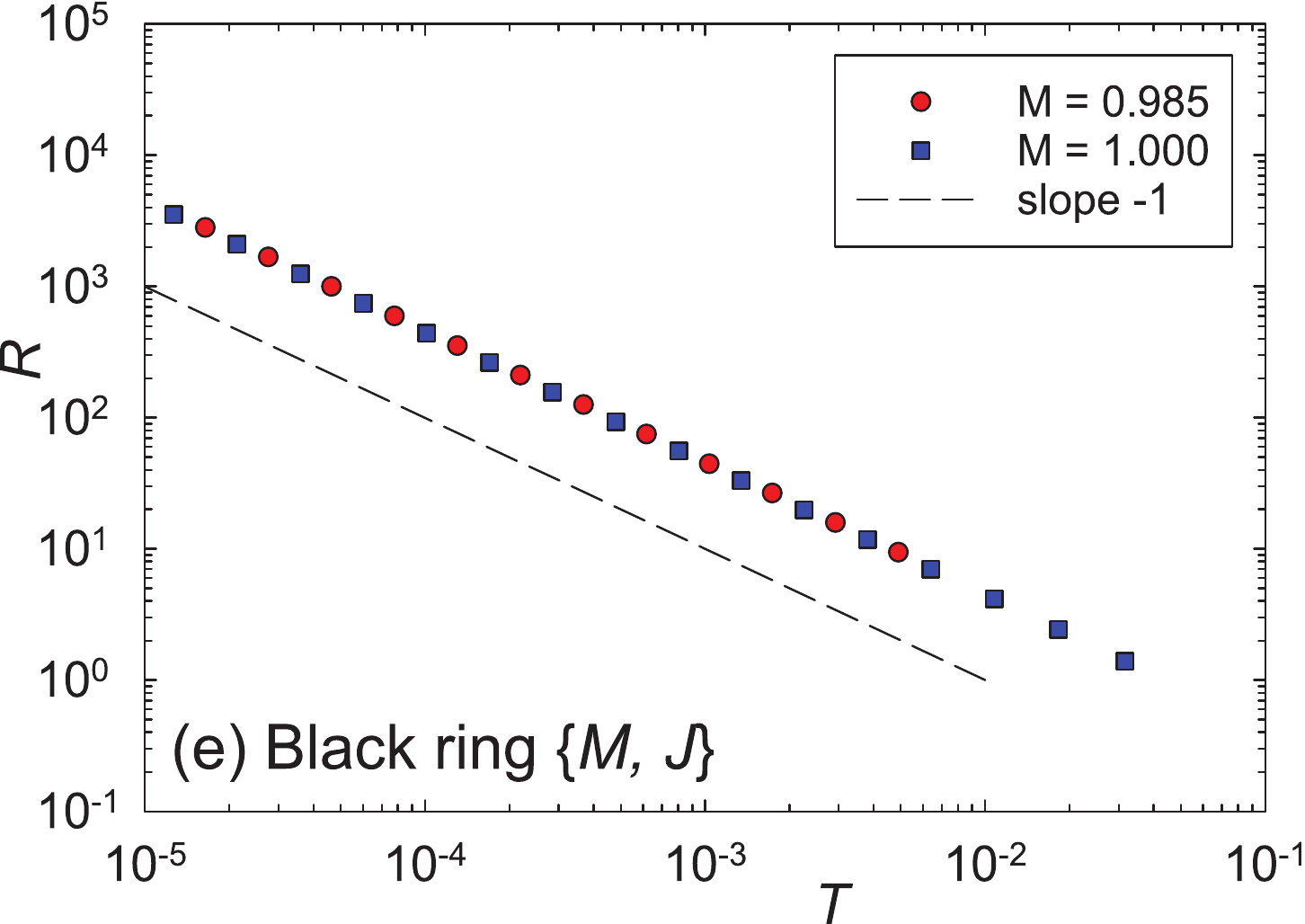} &
    \includegraphics[width = 0.33\linewidth]{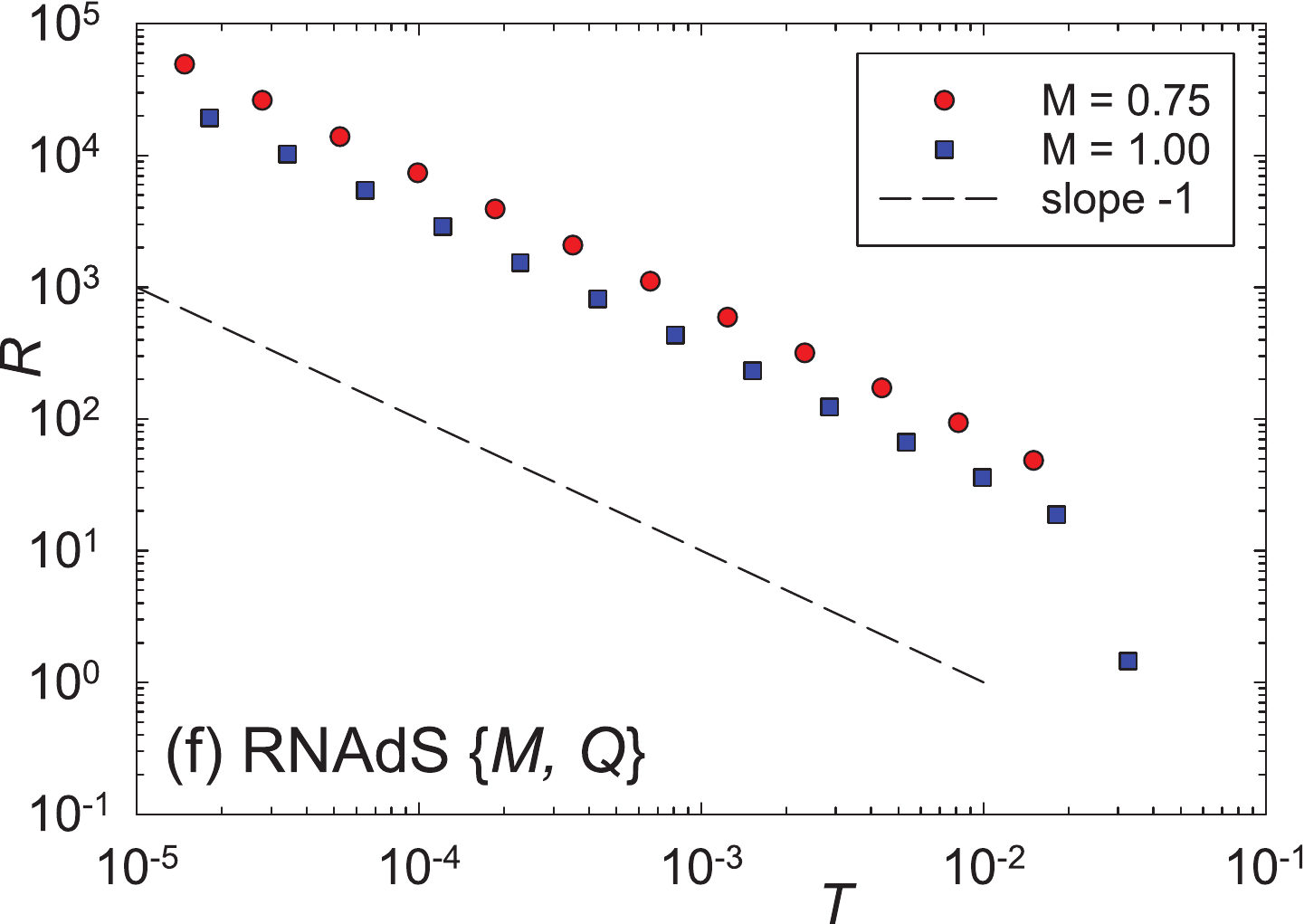}
    \\
    \includegraphics[width = 0.33\linewidth]{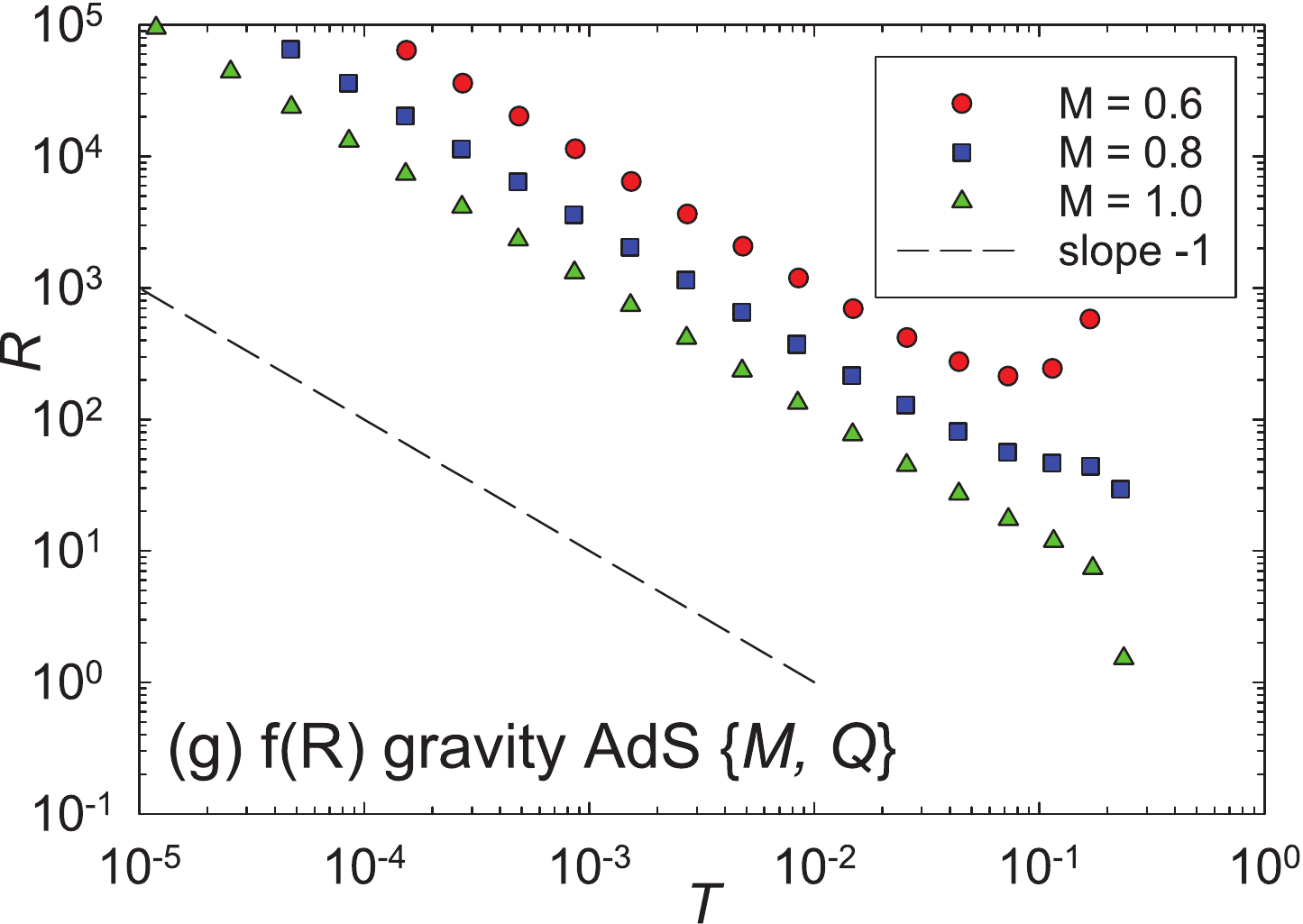}&
    \includegraphics[width = 0.33\linewidth]{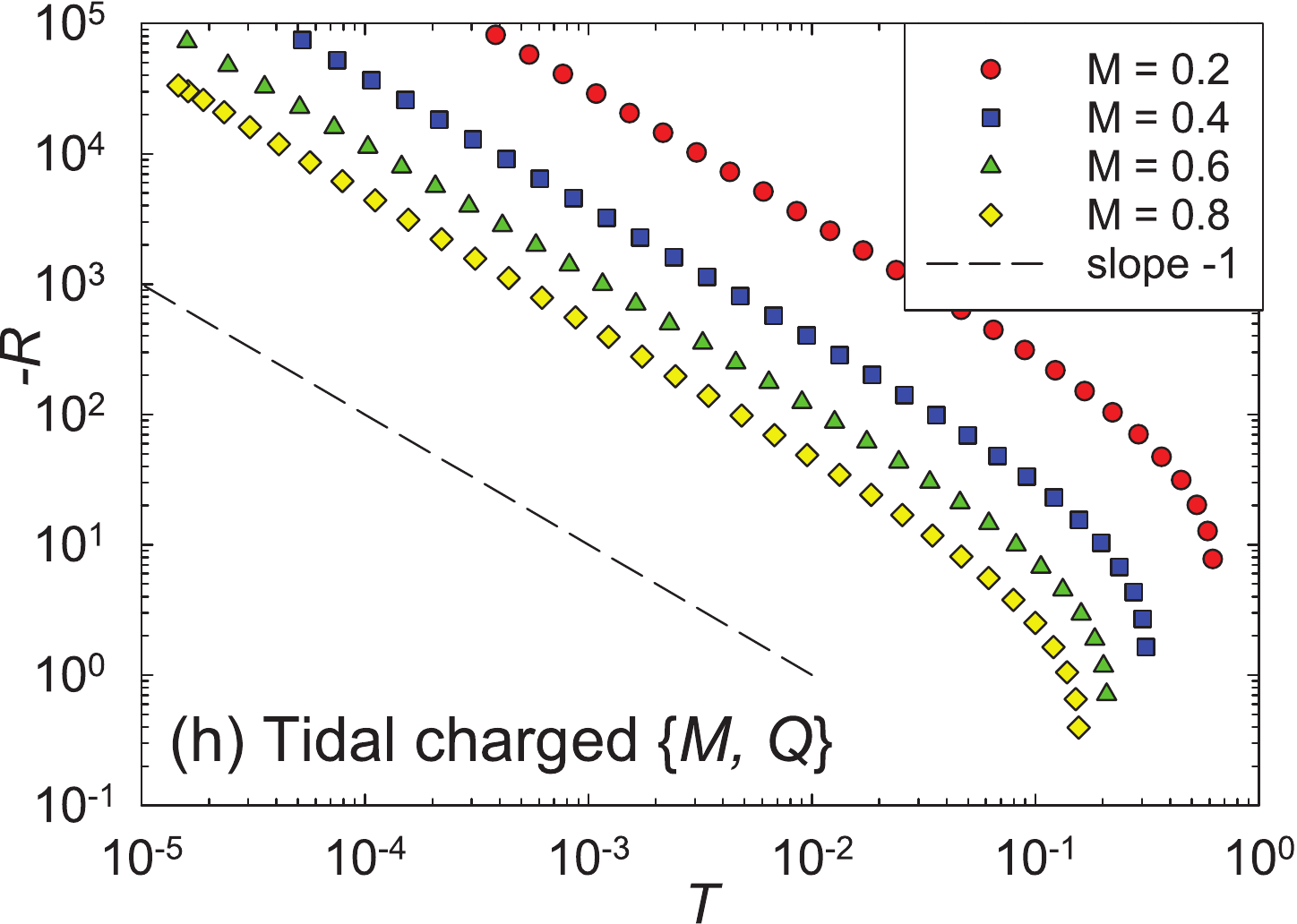} &
    \includegraphics[width = 0.33\linewidth]{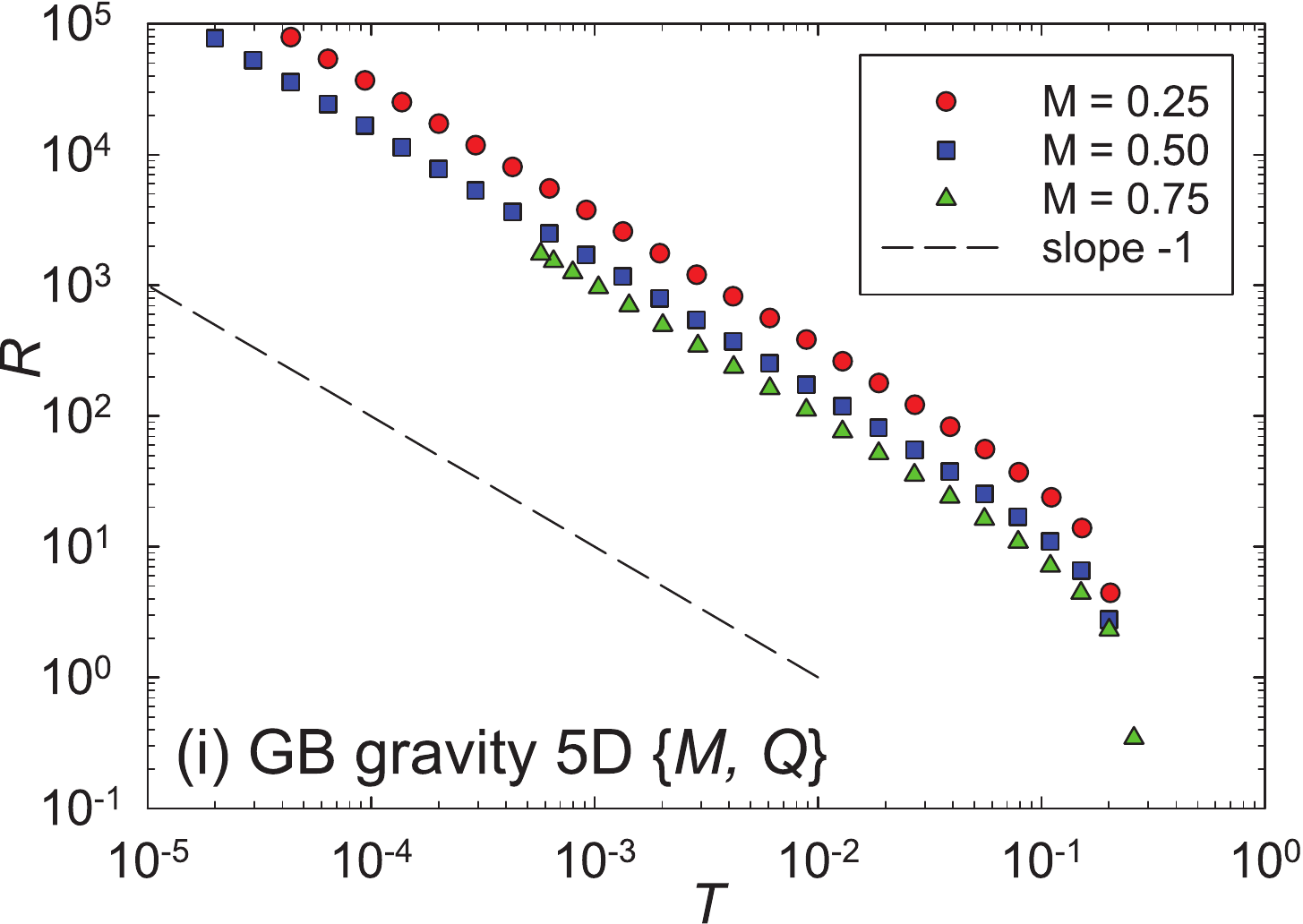}
    \\
    \includegraphics[width = 0.33\linewidth]{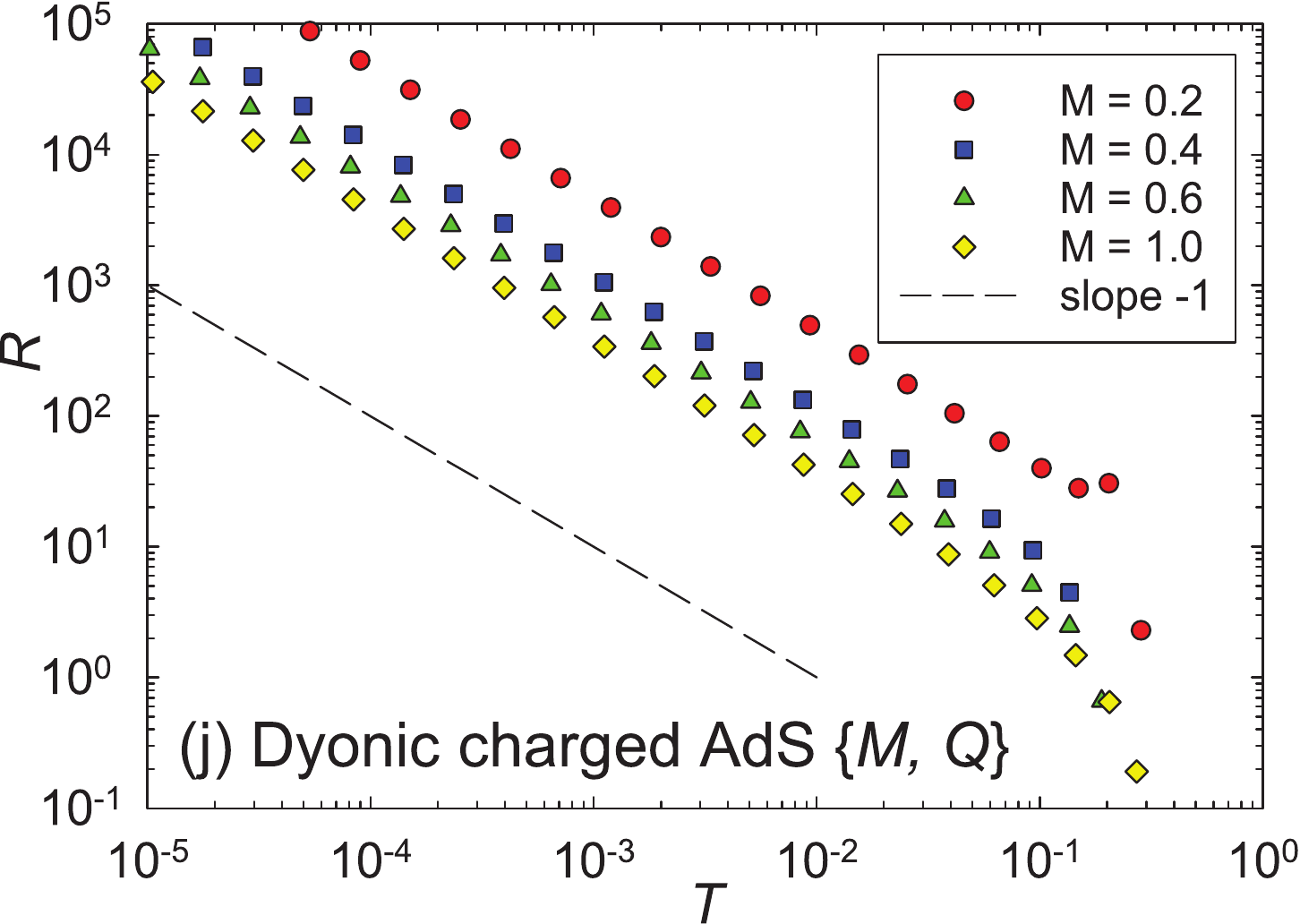} &
    \includegraphics[width = 0.33\linewidth]{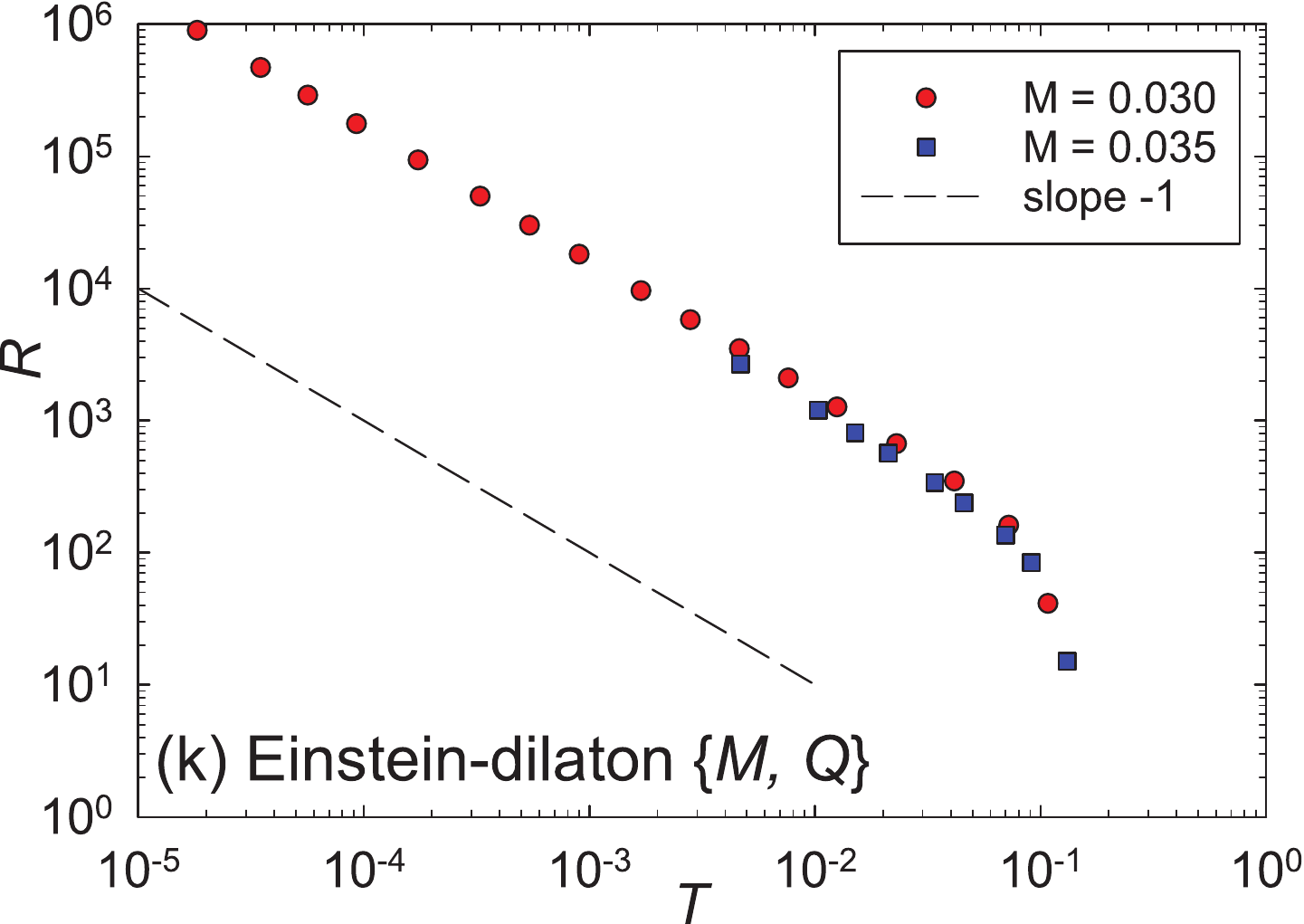} & \includegraphics[width = 0.33\linewidth]{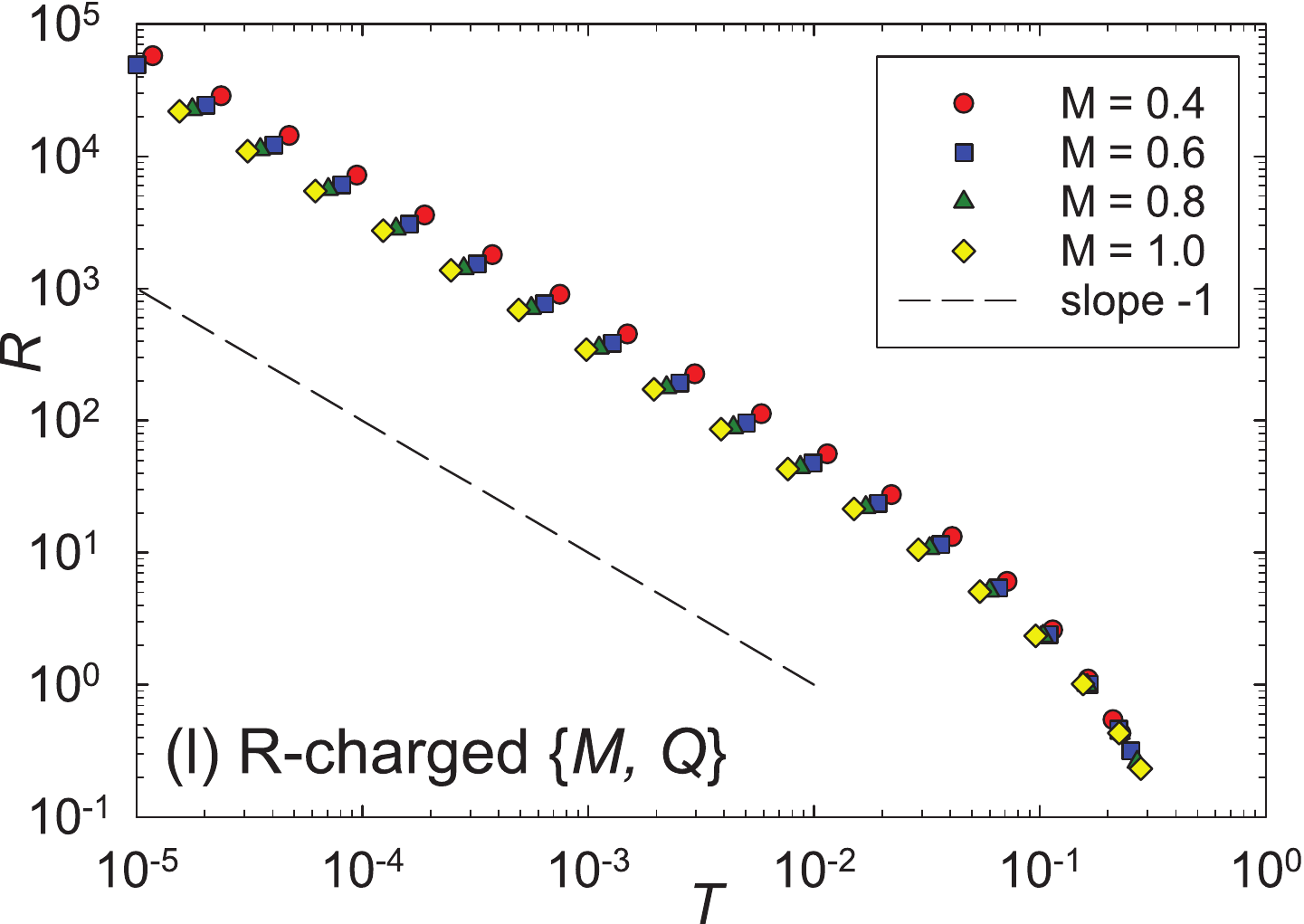}
    \\
    \end{tabular}
    \caption{Fits to $R$ data versus $T$ for various constant values of $M$ near the extremal limit. Note that $R$ for the tidal charged figure (h) is negative.}
    \label{F: Fitted Plots}
    \end{figure}

\begin{table}
\centering
\begin{tabular}{l|c|c||l|c|c}
    Model & Mass $M$ & $\log_{10} c$ & Model & Mass $M$ & $\log_{10} c$\\
\hline
\hline
    Kerr & 0.4 &  0.2907         & $f(R)$ gravity  & 0.6   & 0.9575\\
         & 0.6 & -0.2376         &                                    & 0.7   & 0.6684\\
         & 0.8 & -0.6124         &                & 0.8   & 0.4303\\
         & 1.0 & -0.9031         &                & 0.9   & 0.2276\\
         &     &                 &                & 1.0   & 0.0501\\
    \hline
    KN $(M, J)$ & 0.6 & -0.1284  &  Tidal charged    & 0.2   & 1.4948\\
                & 0.8 & -0.5544  &                   & 0.4   & 0.5917\\
                & 0.9 & -0.7207  &                   & 0.6   & 0.0635\\
                & 1.0 & -0.8669  &                   & 0.8   & -0.3114\\
    \hline
    KN $(M, Q)$          & 0.92 & -0.7198 & GB-AdS   & 0.25  & 0.5371\\
                         & 0.94 & -0.7330 &          & 0.50  & 0.1929\\
                         & 0.96 & -0.7467 &          & 0.75  & -0.0028\\
                         & 0.98 & -0.7608 &          &       &        \\
                         & 1.00 & -0.7753 &          &       &        \\
    \hline
    Kerr 5D              & 0.85 & -0.4097            & Dyonic-AdS & 0.2   & 0.6712\\
                         & 0.90 & -0.4594            &            & 0.4   & 0.0698\\
                         & 0.95 & -0.5064            &            & 0.6   & -0.1827\\
                         & 1.00 & -0.5509            &            & 0.8   & -0.3250\\
                         &      &                    &            & 1.0   & -0.4189\\
    \hline
    Black ring          & 0.985 & -1.3360           & E-d         & 0.030 & 1.2111\\
                         & 0.990 & -1.3404           &            & 0.035 & 1.0789\\
                         & 0.995 & -1.3447           &            &       &       \\
                         & 1.000 & -1.3491           &            &       &       \\
    \hline
    RN-AdS               & 0.80  & -0.2122           & $R$-Charged          & 0.3 & -0.0678\\
                         & 0.90  & -0.3427 &         & 0.5 & -0.2457\\
                         & 1.00  & -0.4557 &         & 0.7 & -0.3569\\
                         &       &         &         & 0.9 & -0.4370\\
    \hline
    \end{tabular}
    \caption{The fit parameters $\log_{10} c$ for the twelve models considered here. Each fit is for a data set all with the same value of $M$. In the extremal limit, we expect $R\to c\,T^{-1}$, or the linear function $\log_{10} R=\log_{10}c-\log_{10}T$. For each data set, we fit the linear function to the three data points with the smallest values of $T$. Fits and data are graphed in Figure \ref{F: Fitted Plots}. Each fit had an exponent for $T$ very close to $-1$ (within $0.01\%$) and we do not show their tiny deviations. For tidal charged, we did the logarithmic fit to $-R$, since $R$ is negative.}
  
    \label{fit coefficient table}
    \end{table}

\par
The introduction of a non-zero $Q$ does not affect the essential extremal limiting behavior of $R$ at constant $M$ given in Eq. (\ref{10986}): $R\propto +1/T$. For $Q$ less than about 0.85, the BHT is unstable near the extremal curve \cite{Ruppeiner2007}, and this may diminish its interest. For larger $Q$, there is a region of thermodynamic stability adjacent to the extremal curve. However, highly charged black holes are physically very unlikely to occur.

\par
The temperature dependence of $R$ for Kerr-Newman $\{M, \, J\} (Q=0.4)$ near the extremal curve is shown in Figure \ref{F: Fitted Plots}(b), which shows $R$ as a function of $T$, with $M$ fixed at various values. On the log-log scale, the asymptotic behavior is a straight line with slope $-1$, consistent with $R\propto 1/T$. Values of the fit coefficients are shown in Table \ref{fit coefficient table}. They show a trend visible in all of our fits: at given $T$, $R$ tends to be smaller at larger $M$. The figures in Figure \ref{F: Fitted Plots} are to be compared with the figures in Figure \ref{FigureFermi} for the ideal Fermi gas, which they resemble in their temperature dependence.

\subsection{Kerr-Newman $\{M, \, Q\} (J=0.7)$ \label{SS: KNMQ}}

Consider now the case where $\{M,Q\}$ fluctuate, and $J>0$ is held fixed. ($J=0$ corresponds to the Reissner-Nordstr$\ddot{\mbox{o}}$m solution to GR, which is known to have $R=0$ \cite{Aaman2003,cai1999}). The functions $S$ and $T$ for this model are the same as for the Kerr-Newman $\{M, \, J\}$ model, Eqs. (\ref{KNentropy}) and (\ref{KNTemp}). $R$ is given in \cite{Ruppeiner2008} with a slightly different scaling from here:

\begin{equation}R=-\frac{\begin{array}{lll} (L-1) (L+1) (K^5 L^2+8 K^5-4 K^4 L^2+4 K^4+4 K^3L^4+\\\qquad 14 K^3 L^2-36 K^3-8 K^2L^4+40 K^2L^2-32 K^2+3 K L^6-\\\qquad\qquad 6 K L^4-36 K L^2+48 K-4 L^6+36 L^4-96 L^2+64)\end{array}}{2 K
   M^2 \left(K^4-K^3 L^2+4 K^3+K^2 L^2+2 K^2-K L^4-2 K L^2+4 K+2 L^4-10 L^2+8\right)^2}.\end{equation}

\noindent Near the extremal curve, BHT \cite{Ruppeiner2007} is thermodynamically stable for all states. Increasing $T$ from the extremal curve along a line of constant $Q/M$, has us encounter a line of phase transitions along which $R\to+\infty$; see Figure 6 of \cite{Ruppeiner2007}. But this line is not particularly interesting in this study, so we do not pursue it. The extremal limiting $R$ is given by

\begin{equation}
    R \to \frac{1}{4 L^2 M^3 T},
    \label{45620}
\end{equation}

\noindent exactly the same as for Kerr-Newman $\{M, \, J\}$, Eq. (\ref{10986}). Again, we analytically find $R\propto +1/T$ in the extremal limit along lines of constant $M$.

\par
Figure \ref{F: 3D Plots}(c) shows a contour plot of $R$ for Kerr-Newman $\{M, \, Q\}$ with fixed $\{J,\Lambda\}=\{0.7,0\}$. Clearly, $R$ increases with decreasing $T$. The line of phase transitions is not visible in this representation. The fitted analysis at constant $M$ is shown in Figure \ref{F: Fitted Plots}(c), and clearly $R\propto+1/T$. The fit parameters are listed in Table \ref{fit coefficient table}.

\subsection{Kerr 5D $\{M,\,J\}$}

The Kerr 5D BHT model is more exotic. Myers and Perry \cite{Myers1986} constructed its GR metric by adding a fourth spatial dimension to the Kerr metric. {\AA}man and Pidokrajt \cite{Aaman2006} added $\{M,J\}$ BHT fluctuations, and constructed the thermodynamic geometry. In five dimensions, two angular momenta are possible, but we consider only one.

\par
We start with the thermodynamic equation for the mass $M=M(S,J)$ in general space-time dimension $d$ \cite{Aaman2006}:

\begin{equation}
    M(S,J)=\frac{d-2}{4} S^{(d-3)/(d-2)}\left(1+\frac{4J^2}{S^2}\right)^{1/(d-2)}.
    \label{769299422}
\end{equation}

\noindent Solving this equation for $S$ (pick the largest real root) and setting $d=5$ yields

\begin{equation}
    S(M,J)=\frac{2}{3} \sqrt{\frac{16 M^3}{3}-9 J^2}.
     \label{10057324}
\end{equation}

\noindent The temperature is

\begin{equation}
    T=\left(\frac{9}{32 M^2}\right) S.
\end{equation}

\noindent With the fundamental thermodynamic equation Eq. (\ref{10057324}) we can construct the full thermodynamic geometry for this system, starting with the thermodynamic metric Eq. (\ref{E: thermo metric}). With only one angular momentum, the thermodynamic geometry is a two-dimensional space parameterized by $(M,J)$.

\par
Eq. (\ref{RicciScalar}) for $R$ produces the remarkably simple expression \cite{Aaman2006}:

\begin{equation}
    R = \frac{1}{S}.
\end{equation}

\noindent A minus sign was omitted on the right-hand side because of the different sign convention between \cite{Aaman2006} and us. Clearly,

\begin{equation}
    R = \frac{9}{32 M^2 T},
\end{equation}

\noindent an expressions that shows the familiar $R\propto +1/T$ behavior over the full range of thermodynamic states, and not just in the extremal limit. This result is confirmed numerically, with fixed $\{Q,\Lambda\}=\{0,0\}$, as seen in Figures \ref{F: 3D Plots}(d) and \ref{F: Fitted Plots}(d). The fit coefficients are shown in Table \ref{fit coefficient table}. Clearly even BHT models in 5D spacetimes show a $R\propto +1/T$ divergence in the extremal limit.

\subsection{Black ring $\{M,\,J\}$}

Even more exotic is the black ring (BR) model \cite{Arcioni2005}. General BR models can be described by a GR line element written in terms of the standard parameters $\{M, J, Q, \Lambda\}$. Following Ref. \cite{Arcioni2005} we consider only two-dimensional thermodynamic geometries of BR systems, with fluctuating $\{M,J\}$. We fix $\{Q,\Lambda\}=\{0,0\}$, corresponding to an uncharged, asymptotically flat space. 

\par
Define first the parameters $\mu=8M/(3\pi)$, $a=3J/(2M)$, and $x=a/\sqrt{\mu}$. For some pairs of $\{M,J\}$ values, there may be two values of $S$, corresponding to a large and a small black hole. We worked out the extremal limit for the small black hole, for which we have the parameter

\begin{equation}
    \nu = \sqrt{8}x \Bigg(\frac{1}{\sqrt{3}}\cos \Theta + \sin \Theta \Bigg) - 1,
\end{equation}

\noindent where 

\begin{equation}
  \Theta = \frac{1}{3} \arctan\sqrt{\frac{32}{27}x^2 - 1}. 
\end{equation}

\noindent It was reported that the small black hole is nowhere thermodynamically stable, but that the large black hole has regimes of stability \cite{Arcioni2005}. In this sense it might have been better to work out the large black hole case first, but the small one gives interesting results also.

\par
The entropy and temperature are \cite{Arcioni2005}:

\begin{equation}
    S = \displaystyle{\frac{\sqrt{2}}{4}\,\pi^2 \mu^{3/2}\sqrt{\nu(1-\nu)}},
\end{equation}

\noindent and

\begin{equation}
    T = \frac{1}{2\sqrt{2}\,\pi \sqrt{\mu}}\sqrt{\frac{1-\nu}{\nu}}.
\end{equation}

\par
The thermodynamic scalar curvature is

\begin{align}
    R = \frac{2\sqrt{2}}{\pi^2 \mu^{3/2}} \frac{\sqrt{\nu}(\nu^2 +2\nu -2)}{(2\nu -1)^2\sqrt{1 - \nu}}.
\end{align}

\noindent The exact analytic temperature dependence of the scalar curvature isn't immediately manifest, so we turn to numerical methods. Figure \ref{F: 3D Plots}(e) shows the familiar, positive, asymptotic behavior in $R$ with decreasing $T$ and fixed $\{Q,\Lambda \}=\{0,0\}$. Fit plots at constant $M$ are shown in Figure \ref{F: Fitted Plots}(e), and they show the extremal divergence $R\propto +1/T$. Fit values are shown in Table \ref{fit coefficient table}.

\subsection{Reissner-Nordstr{\"om} AdS $\{M,\,Q\}$} \label{S: RN AdS}

In this section we investigate the thermodynamic scalar curvature $R$ of the Reissner-Nordstr\"om black hole in AdS space, where we have a cosmological constant $\Lambda$ not zero.

\par
We take the cosmological constant $\Lambda$ to be strictly negative and parameterized by,

\begin{equation}
    \Lambda = -\frac{(d-1)(d-2)}{2\,l^2}, \label{30986745}
\end{equation}

\noindent where $d$ is the spacetime dimension, and $l$ is the AdS length parameter. For $d=4$, we have:

\begin{equation}
    \Lambda = -\frac{3}{l^2}. \label{6578439}
\end{equation}

\noindent A non-zero $\Lambda$ gives the Reissner-Nordstr\"om black holes a non-zero $R$. $\Lambda=0$ has the trivial $R=0$, as mentioned earlier.

\par The entropy $S(M, J, Q, \Lambda)$ is too cumbersome to write here. However, \AA man et al. \cite{Aaman2003} give a compact expression relating the mass and the entropy:

\begin{equation}
M = \frac{\sqrt{S}}{2}\left( 1 + \frac{S}{l^2} + \frac{Q^2}{S}  \right).
\end{equation}

\noindent This equation allows an easy calculation of $R$ using the Weinhold metric method in Eq. (\ref{145987}). The temperature is,

\begin{equation}
    T = \frac{1}{4\sqrt{S}} \bigg( 1 + \frac{3S}{l^2} - \frac{Q^2}{S} \bigg).
\end{equation}

\noindent Finally, the thermodynamic scalar curvature of such a system is given by \cite{Aaman2003}

\begin{equation}
    R = \frac{9}{l^2}\frac{\big( \frac{3S}{l^2} + \frac{Q^2}{S} \big) \big( 1 - \frac{S}{l^2} - \frac{Q^2}{S} \big)}{\big(1 - \frac{3S}{l^2} - \frac{Q^2}{S} \big) \big( 1 + \frac{3S}{l^2} - \frac{Q^2}{S}  \big)}.
\end{equation}

\noindent While this is the first model that we have considered not embedded in asymptotically flat space, Figures \ref{F: 3D Plots}(f) and \ref{F: Fitted Plots}(f) show a similar extremal limiting behavior as the previous models: $R\propto +1/T$. In the figures, we worked with fixed $\{J,\Lambda\}=\{0,-0.1\}$. $l$ is given by Eq. (\ref{6578439}). Fit values of our coefficients are shown in Table \ref{fit coefficient table}.

\subsection{$f(R)$ gravity AdS $\{M,\,Q\}$}

In this section we analyze an instance of $f(R)$ gravity \cite{Capozziello2011, li2016,Moon2011}. This class of theories generalizes the dependency of the relativistic Ricci scalar curvature $R$ in the Einstein-Hilbert action:

\begin{equation}
    I = \int f(R)\sqrt{-g} \, d^4x,
\end{equation}

\noindent where $f(R)$ is some function of $R$. We recover the standard GR when $f(R) = R$. The specific model considered here is a charged AdS black hole in $f(R)$ gravity with constant curvature $R_0$ \cite{li2016}.

\par
The GR line element is

\begin{equation}
    ds^2 = -N(r)dt^2 + \frac{dr^2}{N(r)} + r^2 \left(d\theta^2 + \sin{\theta}^2 d\phi^2 \right),
\end{equation}

\noindent where

\begin{equation}
    N(r) = 1 - \frac{2m}{r} + \frac{q^2}{b\,r^2} - \frac{R_0}{12}\,r^2
\end{equation}

\noindent is the discriminant function. Here, the constant $b=1+f'(R_0)$, and $m$ and $q$ are related to mass and charge via $M=m\,b$ and $Q=q/\sqrt{b}$. We take $b>0$. This black hole solution reduces to the RN-AdS black hole when $b=1$ and $R_0=-12/l^2$. We follow the lead of Li and Mo \cite{li2016}, and set $b=1.5$ and $R_0=-12$. The later equality has $\Lambda=-3$, given by Eq. (\ref{6578439}) above.

\par
The entropy is

\begin{equation}
    S=\pi\,b\,r_+^2,
\end{equation}

\noindent where $r_+^2$ is the radius of the event horizon. We find $r_+^2$ from $S$, and we find $S$ by solving the mass equation (pick the largest real root)

\begin{equation}
    M = \frac{12 b^2 \pi^2 Q^2 + 12 b\,\pi S - R_0\,S^2}{24\pi^{3/2} \sqrt{bS} }.
\end{equation}

\noindent The temperature is

\begin{equation}
    T = \frac{1}{4\pi r_+}\left(1 - \frac{q^2}{b\,r_+^2} - \frac{R_0 r_+^2}{4}\right).
\end{equation}

\par 
The analytic expression for the scalar curvature follows directly from Eq. (36) of ref. \cite{li2016}:

\begin{equation}
    R = \frac{A(S,Q)}{(4 b^2 \pi^2 Q^2 - 4 b \pi S -R_0 S^2)^3(4 b^2 \pi^2 Q^2 -4 b \pi S + R_0 S^2)},
\end{equation}

\noindent where the numerator is

\begin{align}
    A(S, Q) = &-1280 b^7 \pi^7 Q^6 + 64 b^6 \pi ^6 Q^4 (8 - 7 Q^2 R_0)S + 128 b^5 \pi^5 Q^2 (6 + Q^2R_0)S^2 \nonumber \\
    & + 16 b^4 \pi ^4 Q^2 R_0 (20 - 3 Q^2 R_0) S^3 -336 b ^3 \pi ^3 Q^2 R_0^2 S^4 + 4 b^2 \pi^2 R_0^2(4 - 9 Q^2 R_0)S^5  \\ \nonumber 
    & + 16 b \pi R_0^3 S^6 + 3 R_0^4 S^7.
\end{align}

\noindent Figure \ref{F: 3D Plots}(g) shows the characteristic asymptotic behavior of the thermodynamic curvature $R$ at low temperatures. We worked with fixed $\{J,\Lambda\}=\{0,-3\}$. The results of the fitting algorithm, shown in Figure \ref{F: Fitted Plots}(g), confirms that $R \propto +1/T$ in the extremal limit. Fit values of our coefficients are shown in Table \ref{fit coefficient table}.

\subsection{Tidal charged $\{M,\,Q\}$}

The tidal charged black hole model \cite{dadhich2000, Pido2011} comes from string theory, and is also analyzable by our methods. The mass $M$ may be written as a function of the entropy $S$ and the tidal charge $Q$ \cite{Pido2011}:

\begin{equation}
    M=\frac{\sqrt{S}}{2}\left(1+\frac{Q}{S}\right).
\end{equation}

\noindent Finding the largest real root for $S$ of this equation yields

\begin{equation}
    S(M, Q) = \left(M + \sqrt{M^2 - Q}\right)^2.
\end{equation}

\noindent The temperature follows from our Eq. (\ref{E: Temperature definition}), and agrees with Eq. (13) of \cite{Pido2011}:

\begin{equation}
    T(M, Q) = \frac{\sqrt{M^2 - Q}}{2\left(M + \sqrt{M^2 - Q}\right)^2}.
    \label{1954304}
\end{equation}

\noindent When $Q$ is positive, the tidal charge is related to the electric charge $Q_E$ by $Q = Q_E^2$. In the more general brane-world theories, $Q$ may take on negative values as well, but we consider only positive $Q$ since there is no extremal limit for negative $Q$. It was shown that the BHT for this model is stable regardless the sign of $Q$. But in our analysis, we considered only positive $Q$, since negative $Q$ has no extremal limit.

\par
With our sign convention the thermodynamic scalar curvature is the simple \cite{Pido2011}:

\begin{equation}
    R = -\frac{1}{2\sqrt{M^2 - Q}\,\left(M + \sqrt{M^2 - Q}\right)}.
    \label{986765327}
\end{equation}

\noindent The analytic temperature dependence of $R$ is clear from Eqs. (\ref{1954304}) and (\ref{986765327}):

\begin{equation}
    R = -\frac{1}{4\,T \left(M + \sqrt{M^2 - Q}\right)^3}.
    \label{760287534}
\end{equation}

\noindent In the extremal limit, $Q \to M^2$, and the limiting expression for $R$ is

\begin{equation}
    R \to -\frac{1}{4 M^3 T}.
    \label{6580923}
\end{equation}

\noindent This expression resembles the KN limiting expressions. $R$ in Eq. (\ref{760287534}) is finite except in the extremal limit $T\to 0$, so there are no non-zero $T$ phase transitions.

\par
The numerical analysis of these equations, with fixed $\{J,\Lambda\}=\{0,0\}$, produces similar results to before, but with one essential difference: as seen in Eq. (\ref{6580923}), $R$ diverges negative in the extremal limit for the tidal charged black hole. This divergence is bosonic and not fermionic, marking this model as an anomaly, for which we offer no explanation. Figures \ref{F: 3D Plots}(h) and \ref{F: Fitted Plots}(h) shows the extremal limit divergence, $R\propto-1/T$ for constant $M$. Fit values of our coefficients are shown in Table \ref{fit coefficient table}.

\subsection{Gauss-Bonnet AdS $\{M,\,Q\}$}

\par
Gauss-Bonnet gravity theories are based on a truncation of the Lovelock Lagrangian \cite{lovelock1971} to just terms quadratic in the GR curvature tensor. Sahay and Jha \cite{Sahay2017} worked out a class of such theories with an Einstein-Maxwell framework in 5D AdS space, and the Lagrangian

\begin{equation}
   \mathcal{L} = \frac{1}{16\pi G_d}\Big[R - 2\Lambda + \alpha(R^2 - 4R_{\mu \nu}R^{\mu \nu} + R_{\mu \nu \rho \sigma}R^{\mu \nu \rho \sigma}) \Big] - \frac{1}{4}F_{\mu \nu}F^{\mu \nu},
\end{equation}

\noindent where the $d$-dimensional gravitational constant $G_d$ gets set to unity, $\alpha$ is a coupling constant subject to the constraint

\begin{equation}
    0 \leq \frac{\alpha}{l^2} \leq \frac{1}{8}
\end{equation}

\noindent for $d=5$, the only case considered here, and $F_{\mu\nu}$ denotes the matter content via the gauge field stress tensor.

\par
Varying the Einstein-Hilbert action yields the following GR metric:

\begin{equation}
    ds^2=-f(r)dt^2+\frac{1}{f(r)}dr^2 + r^2 h_{ij}dx^i dx^j,
\end{equation}

\noindent where $f(r)$ is given in \cite{Sahay2017}, along with the gauge field. The $h_{ij}$ are the metric elements of the maximally symmetric Einstein space with constant curvature $(d-2)(d-3)k$. The curvature parameter $k$ was taken to be $+1$, and $\alpha=0.01$.

\par
For $d=5$, the authors \cite{Sahay2017} provide compact formulas for the mass, entropy, and temperature (setting the AdS length parameter $l=1$):

\begin{equation}
    M=\frac{\pi(Q^2+12r^4+12 r^6+24r^2\alpha)}{32r^2},
    \label{385692}
\end{equation}

\begin{equation}
    S = \frac{1}{2}\pi^2r(r^2 + 12\alpha),
\end{equation}

\noindent and

\begin{equation}
    T = \frac{-Q^2 + 12r^4 + 24r^6}{24\pi r^3(r^2 + 4\alpha)}.
\end{equation}

\noindent Our numerical solution method is to solve for the outer event horizon radius $r$ with Eq. (\ref{385692}) for given ${M,Q,\alpha}$ (pick the largest real root). This yields $S$ and $T$, and also the scalar curvature $R$, whose analytic expression is too lengthy to display here, but it is given in \cite{Sahay2017}.

\par
Proceeding numerically (with fixed $\{J,\Lambda\}=\{0,-6\}$), the relevant contour plot is found in Figure \ref{F: 3D Plots}(i). The diverging asymptotic behavior in the extremal limit is clearly present here. Figure \ref{F: Fitted Plots}(i) shows in more detail that the thermodynamic scalar curvature $R$ obeys the same extremal limiting behavior at constant $M$ as our other cases: $R \propto +1/T$. Fit values of our coefficients are shown in Table \ref{fit coefficient table}.

\subsection{Dyonic charged AdS $\{M,\,Q\}$} \label{S: Dyonic Charge}

Dyonic charged AdS black holes characterize solutions to Einstein-Maxwell theories in AdS space, with both an electric charge $q_E$ and a magnetic charge $q_M$ considered. We follow the analysis of \cite{Sengupta2017, Lu2013}, based on static, spherically symmetric black holes. To restrict the thermodynamics to two fluctuating variables, the authors \cite{Sengupta2017} allowed $M$ and $q_E$ to fluctuate at fixed $q_M$. This model has two charges rather that the standard charge $Q$ and angular momentum $J$ used elsewhere in this paper. We handle this formally by letting $Q=q_E$ and $J=q_M$. These black holes thus correspond here to $\{M,Q\}$ fluctuating at fixed $\{J,\Lambda\}=\{0.13,-3\}$.

\par
Dyonic black holes have the space-time metric,

\begin{equation}
    ds^2 = -f(r) dt^2 + \frac{dr^2}{f(r)} + r^2 d\theta^2 + r^2 \sin^2\theta\,d\phi^2.
\end{equation}

\noindent The lapse function $f(r)$ is given by

\begin{equation}
    f(r) = 1 + \frac{r^2}{l^2} - \frac{2M}{r} + \frac{q^2_E + q^2_M}{r^2},
\end{equation}

\noindent where $l$ is the AdS length scale that the authors set to unity, corresponding to our $\Lambda=-3$, by Eq. (\ref{6578439}).

\par The spherical symmetry results in an entropy proportional to the square of the outer event horizon radius (found by solving $f(r) = 0$ for the largest real root). The entropy as a function of mass is unwieldy, but we have instead the compact inverse relationship

\begin{equation}
    M = \frac{1}{2}\sqrt{\frac{\pi}{S}}\left(\frac{S^2}{\pi^2} + \frac{S}{\pi} + q_e^2 + q_m^2 \right)
\end{equation}

\noindent to calculate the thermodynamic metric via the Weinhold metric Eq. (\ref{145987}). The temperature is \cite{Sengupta2017},

\begin{equation}
    T = \frac{1}{4 (\pi S)^{3/2} } \left[ 3S^2 + \pi S - \pi^2(q^2_E + q^2_M) \right].
\end{equation}

\noindent The scalar curvature is 

\begin{equation}
    R = \frac{(\pi^2 q_m^2 + 3S^2)\left[ 3\pi^4 (q_e^2 + q_m^2)^2 + \pi^3 S (-3q_e^2 +q_m^2) + 12\pi^2 S^2 (q_e^2 + 3q_m^2) - 9 \pi S^3 + 9 S^4 \right]}{S\left[ \pi^2(q_e^2 + 3q_m^2) - \pi S + 3S^2  \right]^2 \left[ -\pi^2 (q_e^2 + q_m^2) + \pi S + 3S^2  \right]}.
\end{equation}

\noindent This model has a line of phase transitions, which does not enter our discussion.

\par
We analyzed the dependence of $R$ on $\{M,T\}$ with our numerical program. The results are presented in Figures \ref{F: 3D Plots}(j) and \ref{F: Fitted Plots}(j). The asymptotic behavior for $R$ in the extremal limit at constant $M$ follows the characteristic relation $R\propto +1/T$. Fit values of our coefficients are shown in Table \ref{fit coefficient table}.

\subsection{Einstein-dilaton $\{M,\,Q\}$} \label{S: Einstein Dilaton}

Considered next is an instance of the Einstein-dilaton family of black hole models. The scalar curvature $R$ for these models was worked out by Zangeneh et al. \cite{zangeneh2017}, who focused on Lifshitz black hole solutions in Einstein-dilaton gravity with Born-Infeld nonlinear electrodynamics. 

\par
The space-time metric used by these authors was 

\begin{align}
    ds^2 = -\frac{r^{2z} f(r)}{l^{2z}}dt^2 + \frac{l^2 dr^2}{r^2 f(r)} + r^2 d\Omega^2_{n-1}.
\end{align}

\noindent The space-time dimension of the system is $n + 1$, with $n=3$ in this case. The exponent $z$ is the dynamical critical exponent.

\par
Two different classes of solutions were discussed by the authors in \cite{zangeneh2017}: $z = n+1$, and $z\neq n+1$. These two cases manifest themselves in the discriminant function $f(r)$ of the space-time metric, given in Eq. (14) of \cite{zangeneh2017}. Since the event horizon radius is determined by solving $f(r) = 0$ for $r$, the two different cases yield distinct expressions for the entropy $S$.

\par
The first case considered in \cite{zangeneh2017} has $z=1$, for which $z\neq n+1$. This case is covered in the authors' section 4. The authors have several other parameters in their model, which we set to $k=1$, $l=1$, $b=1$, and $\beta=1000$. With $l=1$, we have $\Lambda=-3$. $R$ for this model is too lengthy to display here (and so too are the entropy and temperature), so our analysis is purely numerical. The asymptotic behavior $R\propto +1/T$ at constant $M$ seen in previously considered models is visible for this model as well, as seen in Figures \ref{F: 3D Plots}(k) and \ref{F: Fitted Plots}(k), with fixed $\{J,\Lambda\}=\{0,-3\}$. Fit values of our coefficients are shown in Table \ref{fit coefficient table}.

\subsection{$R$-charged $\{M,\,Q\}$}

Our final model corresponds to a black hole arising from gauged supergravity. Sahay et al. \cite{Sahay2010} worked out the BHT thermodynamics for a 5-dimensional $R$-charged black hole, which may have one, two, or three nonzero $R$-charges. Only the case with three nonzero $R$-charges has an extremal limit, so we considered only it. Let $a_i$ denote the charge parameter of the $i$'th $R$-charge, with the index $i$ having values 1,2, or 3. We simplify by setting all of the charge parameters equal to one another: $a_1=a_2=a_3=a$ \cite{Sahay2010}.

\par
The space-time metric for such a black hole is \cite{Sahay2010,behrndt1999},

\begin{equation}
    ds^2 = -(H_1 H_2 H_3)^{-2/3} f(r) dt^2 + (H_1 H_2 H_3)^{1/3}(f(r)^{-1} dr^2 + r^2 d\Omega_{3,k}),
\end{equation}

\noindent where the $H$ factors are related to the charge parameter $a$,

\begin{equation}
    H_i = 1 + \frac{a_i}{r^2},
\end{equation}

\noindent and the discriminant $f(r)$ is defined by,

\begin{equation}
    f(r) = k - \frac{\mu}{r^2} + \frac{r^2}{l^2}H_1 H_2 H_3,
\end{equation}

\noindent with $\mu$ the mass parameter. In 5D we have the space-time coordinates $(t,r,\psi,\theta,\phi)$. Finally, $d\Omega_{3,k}$ is the angular volume element. We work here only with $k=0$, for which the angular volume element is

\begin{equation}
d\Omega_{3,0} = d\psi^2 + \psi^2(d\theta^2 + \sin{\theta}^2 d\phi^2 ).
\end{equation}

\par
The event horizon radius $r_+$ is found in the standard way, by solving for the largest real positive root to $f(r) = 0$. We calculate numerically for given $k$, $\mu$, $a$, and $l$. The mass $M$ and the charge $Q$ are given by $M=3\mu/2+3a$ and $Q=[a(\mu+a)]^{1/4}$. For a pair of grid parameters $(M,Q)$ these equations may be used to generate corresponding values of $(\mu,a)$. We set $J=0$ and $l=1$.

\par
The entropy and the temperature of this black hole is found to be \cite{Sahay2010}

\begin{equation}
    S = 2\pi(r_+^2 + a)^{3/2},
\end{equation}

\noindent and

\begin{equation}
    T = \frac{1}{2\pi}\frac{2r_+^2 + a}{\sqrt{r_+^2 + a}}.
\end{equation}

\noindent The scalar curvature is

\begin{equation}
R = \frac{3a - 2r_+^2}{\pi(2r_+^2 + a)(a - 2r_+^2)\sqrt{r_+^2 + a}}.
\end{equation}

\par 
We calculate numerically with these formulas, with fixed $\{J,\Lambda\}=\{0,-3\}$. Results are shown in Figures \ref{F: 3D Plots}(l) and \ref{F: Fitted Plots}(l), which confirm the expected $R\propto +1/T$ result at constant $M$. Fit values of our coefficients are shown in Table \ref{fit coefficient table}.

\section{Proportionality constants}

In this broad study of the extremal BHT limit, an issue that we did not explore in any detail was the constant of proportionality $c$ in $R=c/T$, other than its sign. It was found previously \cite{Ruppeiner2008} in the extremal limit that for Kerr $\{M,J\}$ and Kerr-Newman $\{M,J\}$, $\{M,Q\}$, and $\{J,M\}$, the product of $R$ and the heat capacity $C_{J,Q}$ is unity. The same holds for the 2D ideal Fermi gas at small $T$. This correspondence among proportionality constants between BHT and OT would seem to be another strong indication of a connection between the ideal Fermi gas and BHT microstructures. Thus it might have seemed logical to expand on this theme in this research. But we refrained from doing so for several reasons: 1) heat capacities are generally difficult to calculate for BHT, 2) there is a priori no best choice of heat capacity (there are several possibilities), and 3) none of the heat capacities are thermodynamic invariants, and invariance is a property that we have emphasized here. (For help in evaluating BHT heat capacities, see \cite{mansoori2015}). We leave this interesting issue as a topic for the future, perhaps best examined first for individual models, both ideal Fermi and BHT, and not as part of a broad survey as we undertake here.

\par
One might also investigate $c$ just for $R$ alone as to its dependence on the parameters $\{M,J,Q,\Lambda\}$. The exact results presented in Sections 5.1, 5.2, and 5.3 already give some insight, and they are supplemented by our data tabulation in Table \ref{fit coefficient table}. There is, however, no particular correlation between $c$ in BHT and $c$ in the ideal Fermi gasses presented in Section 3. Perhaps we need to find a new Fermi model within OT that better matches to BHT. Or perhaps the connections between $R$ and the heat capacities mentioned above already points to the adequacy of the present ideal Fermi model in two or three dimensions. But new ideas seem called for, and so we have not muddied the picture with an attempt to analyze in detail the data in Table \ref{fit coefficient table}.

\section{Conclusion}

This paper presents analysis of twelve black hole models in the extremal limit, where the black hole thermodynamic (BHT) temperature $T\to 0$. The extremal limit is a natural target for investigation since in ordinary thermodynamics (OT) the Physics generally simplifies as one approaches absolute zero. Frequently, effects of complex interparticle interactions freeze out, leaving little left other than basic quantum properties, such as ideal Bose or ideal Fermi statistics. Perhaps this freezing out of complexity holds for black hole microstructures as well in the limit $T\to 0$, and an essential property of the constituent black hole particles would reveal itself. Our main result in our research is that eleven out of twelve of the BHT models we looked at have the thermodynamic curvature $R\propto +1/T$ along curves of constant mass $M$. This is a property in common with the ideal Fermi gas.

\par
The BHT models considered here are characterized by a variety of thermodynamic parameters, many not corresponding directly with those that appear in OT. A meaningful comparison between OT and BHT requires a careful selection of reasonable common parameters. We focused first of all on the thermodynamic scalar curvature $R$ because it is a thermodynamic invariant. In OT, $R$ clearly offers a connection between thermodynamics and microstructures. If thermodynamics is general, as we would hope, this connection should extend to BHT unchanged, marking $R$ as an excellent object to employ in probing BH microstructures. In our research we displayed $R$ as a function of the mass $M$ and the temperature $T$, two parameters with common meanings in the OT and the BHT scenarios. This pair of variables are known to be appropriate for the Kerr and the Kerr-Newman examples. Our focus then on the function $R=R(M,T)$ throughout this paper would appear to be well motivated.

\section{Acknowledgements}

The authors thank Jan {\AA}man, Pankaj Chaturvedi, Narit Pidokrajt, Anurag Sahay, Gautam Sengupta, Ahmad Sheykhi, Shao-Wen Wei, and M. Kord Zangeneh for helpful correspondence.

\section{Appendix} \label{s: Appendix}

In this Appendix we detail our numerical analysis procedure.

\par
The BHT models that we considered all have their states specified by between two to four parameters selected from the canonical list $\{M, J, Q, \Lambda\}$, where $M$ is the mass, $J$ is the angular momentum, $Q$ is the electric charge, and $\Lambda$ is the cosmological constant. We allowed exactly two of these parameters (though never $\Lambda$) to fluctuate, with the remaining two parameters fixed. We call any permutation of the symbols $\{M, J, Q, \Lambda\}$ a CQuartet, with elements $\{X^1, X^2, X^3, X^4\}$. We set parameters in the CQuartet not appearing in the BHT model to zero.

\par
Essential in our method are functions from the literature for the entropy $S$ (or mass $M$), the temperature $T$, and the thermodynamic scalar curvature $R$. These functions may all be numerically evaluated knowing the values of the parameters within the CQuartet. We refer to the list of symbols $\{X^1, X^2, X^3, X^4, S, T, R\}$ as the Septuplet. For calculating numbers, all our literature BHT models yield the mapping CQuartet $\to$ Septuplet.
    
\par
We follow the convention that:

\begin{enumerate}
    \itemsep-1em 
    \item $\{X^1,\,X^2\}$ are the fluctuating parameters.\\
    \item $\{X^3,\,X^4\}$ are the fixed parameters.\\
    \item $X^1$ is always $M$.
\end{enumerate}
    
\noindent For example, consider a case with fluctuating $\{M,Q\}$. If we want to analyze how the function $R(M,T)$ varies with $T$ at constant average $M$, we would take $\mbox{CQuartet} = \{M, Q, J, \Lambda \}$.
    
\par
Our first priority in graphing a BHT model is to generate a two-dimensional grid of points $\{X^1,X^2\}$. The grid generation requires specifying minimum and maximum values for both $X^1$ and $X^2$. These limits bracket $p$ numerical values for $X^1$ and $q$ numerical values for $X^2$. Grid values may be spaced linearly or logarithmically. Logarithmic spacing allows us to crowd points closer together as $T\to 0$. Close spacing between points could result in inadequate precision, necessitating extra places of accuracy for reliable computation.

\par
From the CQuartet we generate the CGrid, structured as
    
\begin{equation}\mbox{CGrid}=\{\mbox{CQuartet},\; \{X^3,X^4\},\; \{row_1,row_2,...,row_p\}\},\label{}\end{equation}

\noindent where CQuartet is the list of symbols, and $\{X^3,X^4\}$ are the fixed numerical values. The quantities

\begin{equation}row_i=\{\{X^1_i,X^2_1\},\{X^1_i,X^2_2\},..., \{X^1_i,X^2_q\}\},\label{}\end{equation}

\noindent with $X^1_i$ the numerical value of the $i$'th element in the list of $X^1$ values, and likewise for $X^2_j$. Each $row_i$ has the same value of $X^1$ all the way across, which is convenient for graphing some quantity as a function of $X^2$, holding $X^1$ fixed. The $\{X^1,X^2\}$ grid values, together with the fixed $\{X^3,X^4\}$ values, allow us to numerically determine values for the complete list of Septuplet entries at all the grid points, assuming that we have a BHT model on the scene.

\par
Our main analysis grid is the general grid GGrid, based on the list of symbols GQuartet=$\{ Y^1, Y^2, Y^3, Y^4 \}$ obeying the following rules:

\begin{enumerate}
    \itemsep-1em
    \item $\{ Y^1, Y^2, Y^3, Y^4 \}$ are four distinct symbols selected from Septuplet.\\
    \item $\{Y^1,\,Y^2\}$=$\{X^1,\,X^2\}$.\\
    \item $Y^4$ is the quantity that we want to analyze (here always $R$).\\
    \item $Y^4$ gets graphed and analyzed versus $\{Y^1, Y^3\}$ (here $Y^3$ is always $T$).\\
\end{enumerate}

\noindent We define

\begin{equation}\mbox{GGrid}=\{\mbox{GQuartet},\{Grow_1,Grow_2,...,Grow_p\}\},\label{}\end{equation}

\noindent where

\begin{equation}Grow_i=\{\{Y^1_i,Y^2_1,Y^3(X),Y^4(X)\},\{Y^1_i,Y^2_2,Y^3(X),Y^4(X)\},...,\{Y^1_i,Y^2_q,Y^3(X),Y^4(X)\}\},\label{}\end{equation}

\noindent with $Y^1_i=X^1_i$, and $Y^2_j=X^1_j$. $Y^3(X)$ and $Y^4(X)$ denote the values of $Y^3$ and $Y^4$ at the values of the CQuartet $X$ at the corresponding CGrid point.

\par
There is one more point that we need to appreciate. The construction of the CGrid is based on limits on $X^1$ and $X^2$ that reflect the ad hoc choices of researchers. Since the idea is to operate very near the extremal limit, it is likely that a number of CGrid points will be beyond the extremal limit. We found that in the BHT models here, such unphysical points reveal themselves as having negative or imaginary $T$. Such points were never included in GGrid.

\par
\begin{figure}[h!]
\centering
\includegraphics[width = 0.5\textwidth]{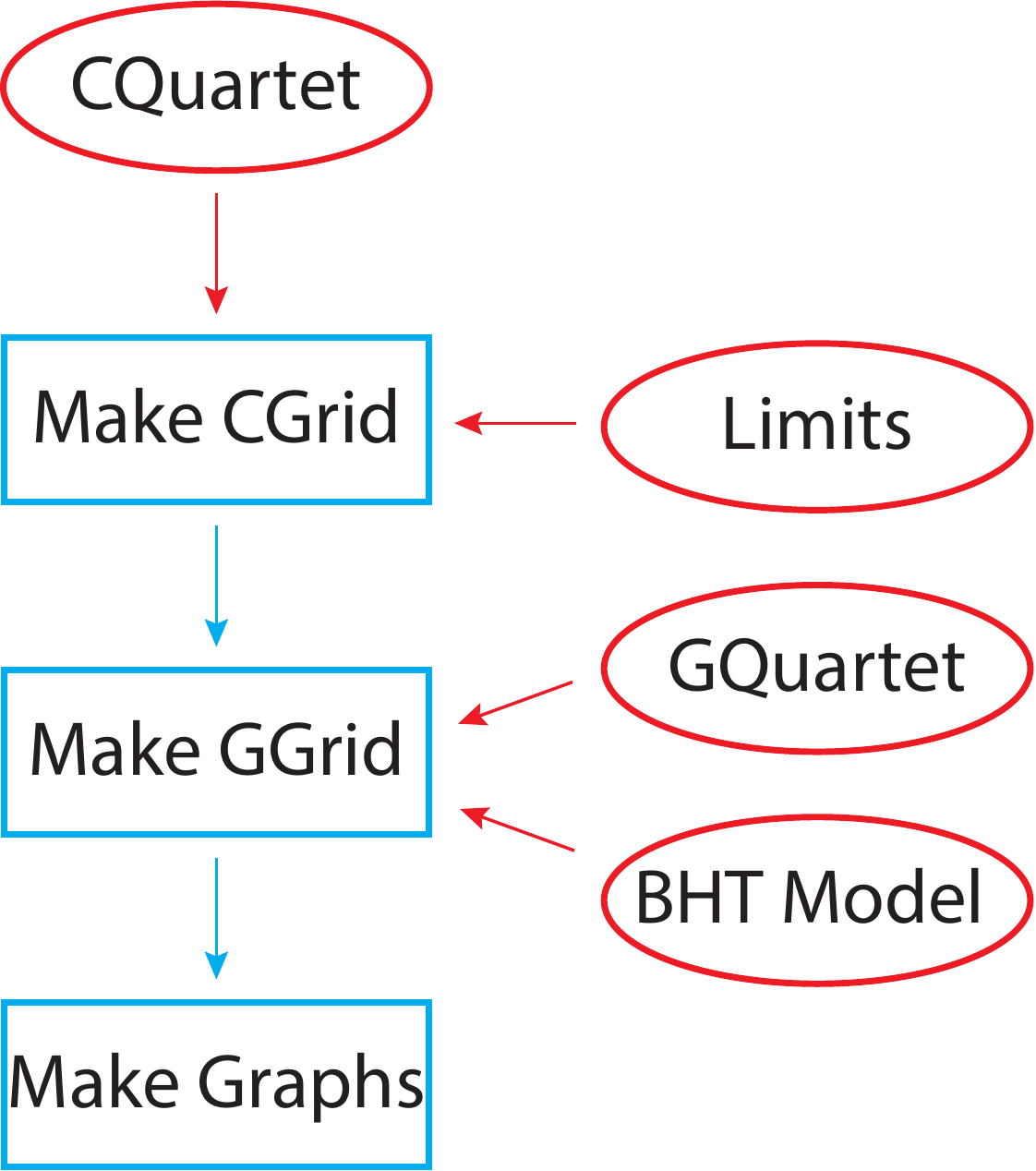}
\caption{The flowchart of our computational algorithm. Red ovals indicate the input needed to get the desired output graphs. This input includes the BHT model consisting of the three functions $S$ (or $M$), $T$, and $R$. The blue rectangles represent the code for generating CGrid, GGrid, and the desired graphs. This code is common to all of our BHT models.}
\label{fig:algorithm}
\end{figure}

Figure \ref{fig:algorithm} shows the broad outline of our computational algorithm.

\bibliographystyle{unsrt}
\bibliography{refs}

\end{document}